\documentclass[reprint, superscriptaddress, nofootinbib, amsmath,amssymb, floatfix, aps, pra]{revtex4-2}

\usepackage{graphicx}
\usepackage{dcolumn}
\usepackage{bm}
\usepackage{hyperref}
\usepackage[mathlines]{lineno}
\usepackage{xcolor}
\usepackage{tabularx}
\usepackage{aas_macros}

\newcommand{\trace}[1]{\mathrm{tr}\left [ #1 \right ]}
\newcommand{\expect}[1]{\left < #1 \right >}
\newcommand{\phihat}{\hat{\phi}}
\newcommand{\phitilde}{\tilde{\phi}}

\newcommand{\Z}{\bm{Z}}
\newcommand{\X}{\bm{X}}

\newcommand{\blockComment}[1]{}

\newcommand{\citeneed}[1]{\textcolor{blue}{CITATION NEEDED}}

\begin{document}

\title{Mapping the Gravitational-wave Background Across the Spectrum with a Next-Generation Anisotropic Per-frequency Optimal Statistic}

\author{Kyle A. Gersbach}
\email{kyle.a.gersbach@vanderbilt.edu}
\affiliation{Department of Physics \& Astronomy, Vanderbilt University,\\
2301 Vanderbilt Place, Nashville, TN 37235, USA}

\author{Stephen R. Taylor}
\affiliation{Department of Physics \& Astronomy, Vanderbilt University,\\
2301 Vanderbilt Place, Nashville, TN 37235, USA}

\author{Bence B\'{e}csy}
\affiliation{
Department of Physics, Oregon State University, Corvallis, OR 97331, USA
}
\affiliation{
Institute for Gravitational Wave Astronomy and School of Physics and Astronomy, University of Birmingham, Edgbaston, Birmingham B15 2TT, UK
}

\author{Anna-Malin Lemke}
\affiliation{II. Institute of Theoretical Physics, Universität Hamburg, Luruper Chaussee 149, 22761, Hamburg, Germany}
\affiliation{Deutsches Elektronen-Synchrotron DESY, Notkestr. 85, 22607 Hamburg, Germany}

\author{Andrea Mitridate}
\affiliation{Deutsches Elektronen-Synchrotron DESY, Notkestr. 85, 22607 Hamburg, Germany}

\author{Nihan Pol}
\affiliation{Department of Physics, Texas Tech University, Box 41051, Lubbock, TX 79409, USA}

\date{September 8, 2025}

\begin{abstract}
With pulsar timing arrays (PTAs) having observed a gravitational wave background (GWB) at nanohertz frequencies, the focus of the field is shifting towards determining and characterizing its origin. While the primary candidate is a population of GW-emitting supermassive black hole binaries (SMBHBs), many other cosmological processes could produce a GWB with similar spectral properties as have been measured. One key argument to help differentiate an SMBHB GWB from a cosmologically sourced one is its level of anisotropy; a GWB sourced by a finite population will likely exhibit greater anisotropy than a cosmological GWB through finite source effects (``shot noise'') and potentially large-scale structure. Current PTA GWB anisotropy detection methods often use the frequentist PTA optimal statistic for its fast estimation of pulsar pair correlations and relatively low computational overhead compared to spatially-correlated Bayesian analyses. However, there are critical limitations with the status quo approach. In this paper, we improve this technique by incorporating three recent advancements: accounting for covariance between pulsar pairwise estimates of correlated GWB power; the per-frequency optimal statistic to dissect the GWB across the spectrum; and constructing null-hypothesis statistical distributions that include cosmic variance. By combining these methods, our new pipeline can localize GWB anisotropies to specific frequencies, through which anisotropy detection prospects---while impacted by cosmic variance---are shown to improve in our simulations from a $p$-value of $\sim0.2$ in a broadband search to $\sim0.01$ in the per-frequency search. Our methods are already incorporated in community-available code and ready to deploy on forthcoming PTA datasets.
\end{abstract}

\maketitle

\section{Introduction} \label{sec:intro}

The first evidence of low-frequency gravitational waves was recently delivered through long-term millisecond pulsar timing array (PTA) campaigns. This cooperative effort from multiple PTA experiments around the world included the North American Nanohertz Observatory for Gravitational waves (NANOGrav) \citep{NG15_evidence}, the European and Indian PTAs (EPTA+InPTA) \citep{EPTA_dr2}, the Parkes PTA (PPTA) \citep{PPTA_dr3}, Meerkat PTA (MPTA) \citep{MPTA_gwb}, and the Chinese PTA (CPTA) \citep{CPTA_dr1}. All of these experiments have shown evidence for distinctive quadrupolar-like correlated timing observations between spatially-separated pulsars, consistent with expectations for a stochastic gravitational wave background (GWB). This evidence ranges from tentative to highly compelling depending on the relative size of the pulsar array and the sensitivity of pulsars therein.

While this observation of a GWB is exciting, we have yet to fully understand its origin \citep{NG15_SMBHB, NG15_newPhys}. The low GW frequency regime to which PTAs are sensitive is expected to primarily comprise the unresolved superposition of signals from a the population of supermassive black hole binaries (SMBHBs) \citep{Becsy2022, Phinney2001}. However, other potential cosmological sources, such as first-order phase transitions, non-standard inflationary scenarios, scalar-induced GWs, and cosmic strings, may also produce a GWB at nanohertz frequencies \citep{NG15_newPhys, Kibble1976}. Spurred by this ambiguity, PTAs continue their efforts to further characterize the GWB and tease out its origin.

Characterization of the GWB is currently focused on the spectral and spatial properties of its intensity distribution, since GWB polarization is beyond current sensitivity \citep{Cruz2024}. Its spectrum can constrain models of SMBHB demographics and binary dynamical evolution, or even be exploited to discriminate multiple overlapping GWBs with different origins \citep{NG15_newPhys, NG15_discrete, Kaiser2022}. On the other hand, the GWB's spatial distribution, as probed through information encoded in inter-pulsar correlations, can constrain the level of signal anisotropy \cite{NG15_anis, Taylor_bumpy, Pol_milestones}. Since cosmological GWB signals are likely to have very minor intensity anisotropies \citep{Olmez2012}, measuring a significant level of anisotropy may indicate an origin for the GWB as a finite population of individual astrophysical sources \citep{Gardiner2024}.

However, most current efforts focus solely on the spectrum, as it is more constrained than the GWB's spatial properties by virtue of pulsar cross-correlations carrying weaker information than the auto-correlations. Also, practically, it is easier to differentiate origin scenarios and ``new physics'' in terms of spectral properites (e.g. \citet{NG15_newPhys}). Nevertheless, some models remain nearly indistinguishable. For these, it may still be possible to comment on the GWB's origin and properties by folding in what spatial information is available, and leveraging such anisotropy search techniques in the future \citep{Mingarelli2013, TaylorGair2013, Mingarelli2017}. Efforts aimed at detecting GWB anisotropies currently employ a frequentist parameter estimation and detection statistic called the PTA optimal statistic (OS) \citep{NG15_anis, MPTA_anis}. The OS provides ready access to pair-wise correlation estimators, with which one can search for anisotropic correlation signatures using a maximum likelihood approach. The OS permits these anisotropic analyses to be both fast to analyze and massively parallelized when compared to Bayesian analyses with equivalent aims \citep{Pol_forcast}.

While powerful, the current implementation of this technique has several limitations. Recent work has shown that the original formulation of the OS does scale to the strong GWB-signal regime \citep{AllenRomano2023, Gersbach2025}. The assumption that pulsar pair correlation estimates are independent breaks down within this regime since it becomes ever more critical to account for the presence of the same pulsar in numerous pair estimates. It has been shown that this assumption is already violated in the latest NANOGrav dataset, and it is likely true for any dataset with a measurable GWB \citep{NG15_pipeline}. 

Another issue arises from the assumption of broadband anisotropy. If a population of SMBHBs is truly the dominant origin of the GWB, this assumption breaks down since each individual binary's GW signal will be slowly evolving and approximately monochromatic over PTA observation baselines \citep{NG15_cw}. Hence each frequency bin is responding to a different sample of the SMBHB population, naturally giving rise to frequency-specific anisotropy, where individual loud sources may dominate, but at different sky locations \citep{Becsy2022}.

Finally, and most crucially, the current OS anisotropy technique does not properly account for cosmic variance \citep{Konstandin2025}. Cosmic variance is a more recently understood issue for PTAs, in which a statistically isotropic GWB can exhibit deviations from the expected spatial correlation pattern \citep{Allen_CV}, even if pulsar and instrument noise were to be completely zero. A statistically isotropic GWB will, when averaged over many realizations of the universe, produce a predictable correlation pattern called the Hellings \& Downs (HD) curve. However, since we only live in one universe, we only see one realization of this GWB. This means that such deviations from HD are still consistent with an isotropic GWB \citep{Konstandin2025}, and must not be confused for anisotropy.

This paper is laid out as follows. We describe the current landscape of frequentist PTA GWB anisotropy searches in \autoref{sec:current}. Next, in \autoref{sec:improve} we discuss our improvements to these techniques, addressing each of the issues outlined above. To test our new improvements, we generate a dataset as described in \autoref{sec:dataset}, after which we outline the specifics of the implementation of these methods in \autoref{sec:implementation}. Our results are shown in \autoref{sec:results}, with a discussion of the key takeaways in \autoref{sec:discussion}. We provide concluding remarks and a discussion of many caveats and key suggestions for further development in \autoref{sec:conclusion}. A busy reader with knowledge of the PTA optimal statistic and the GWB anisotropy formalism may wish to skip to \autoref{sec:improve}.

\section{Existing Frequentist GWB Anisotropy Search Framework} \label{sec:current}

Frequentist methods for detecting and characterizing anisotropy in PTAs use the PTA optimal statistic (OS) \citep{Anholm2009, Chamberlin2015}. This is an optimal-filter cross-correlation statistic designed to both detect, and estimate the properties of, a gravitational wave background (GWB). While originally designed to search for Hellings and Downs (HD) correlations expected of a statistically isotropic GWB \citep{HD}, the OS framework itself is agnostic to the sought-for correlation pattern and therefore can be co-opted as an anisotropic search tool \citep{Sardesai2023}. 

We briefly reiterate the features of the OS that are salient to this present work. For ease of notation, we first define two quantities introduced in \citet{Pol_forcast}:
\begin{equation}
    \X_a = 
    \bm{F}_a^T \bm{P}_a^{-1} \delta t_a
    ,
\end{equation}
and
\begin{equation}
    \Z_a = 
    \bm{F}_a^T \bm{P}_a^{-1} \bm{F}_a
    ,
\end{equation}
where $\delta t_a$ is a vector of pulsar $a$'s timing residuals given by the subtraction of a best-fit timing model's predicted TOAs from observed TOAs; $\bm{F}_a$ is pulsar $a$'s Fourier design matrix; and $\bm{P}_a = \expect{\delta t_a \delta t_a^T}$ is pulsar $a$'s auto-covariance matrix. With these defined, the OS estimator for correlated power between pulsars $a$ and $b$ and its corresponding variance is
\begin{equation} 
    \rho_{ab} =  
    \frac{ \X^T_a \phihat \X_b }
    {\trace{ \Z_b \phihat \Z_a \phihat }}
    ,
    \label{eq:OS_rho_ab}
\end{equation}
\begin{equation} 
    \sigma_{ab,0}^2 = 
    \trace{\Z_b \phihat \Z_a \phihat}^{-1}
    ,
    \label{eq:OS_sig_ab}
\end{equation}
where $\phihat$ represents a unit-amplitude diagonal Fourier-domain covariance matrix, encoding the spectral shape of the GWB in timing residual space \citep{Pol_forcast, Gersbach2025, Taylor2021}. 

To construct these pairwise estimators, we need some way to estimate the total pulsar auto-covariance $P_{a}$ for each pulsar while also informing our spectral shape for $\phihat$. \citet{Vigeland_NMOS} showed that we can leverage the results of a fast Bayesian common uncorrelated red noise (CURN) process analysis to estimate these parameters. Through Markov chain Monte Carlo exploration, we can collect samples from the joint posterior probability distribution of the CURN and pulsar noise parameter space. We then randomly draw from these samples to compute the necessary quantities in the OS, effectively marginalizing over the noise processes of the PTA; hence, this is referred to as the noise-marginalized OS (NMOS). This drastically increases the overall accuracy of the OS and better accounts for the total spread in estimator values \cite{Vigeland2018, Gersbach2025}. One important factor to remember however, is that the spectral shape encoded in $\phihat$ is a fixed quantity and is informed by the initial CURN analysis \citep{Gersbach2025}; often set such that this becomes a diagonal matrix representing a unit-power-law power spectral density (PSD) 
\begin{equation}
    \hat{\phi}_{nn'} = \delta_{nn'} S(f_n) \Delta f = 
    \frac{1}{12 \pi^2 T_{\rm span}}
    \left( \frac{f_n}{f_{\rm yr}} \right) ^{-\gamma} 
    f_{\mathrm{yr}}^{-3}
    ,
\end{equation}
where $n$ and $n'$ index frequency bins; $S(f_n)$ is the PSD of the pulsar timing residuals; $T_{\rm span}$ is the total time-span of the PTA in seconds; and $\gamma$ is the spectral index of the power-law \citep{Chamberlin2015, Gersbach2025}. This unit-amplitude PSD can then be multiplied by an amplitude $A_{\rm gw}^2$ which represents the PSD at frequency $f_{\rm yr} = 1/{\rm yr}$.

We next define a model for the pairwise correlated amplitudes. The OS normalizes these estimators such that, on average, they represent the product of the overlap reduction function (ORF) $\Gamma_{ab}$, and the amplitude of the GWB $A_{\rm gw}^2$, such that $\expect{\rho_{ab}} = \Gamma_{ab} A_{\rm gw}^2$ \citep{Gersbach2025}. For the isotropic case, $\Gamma_{ab}$ represents the HD function evaluated between two pulsars. In the anisotropic case, we must adopt a general ORF, $\Gamma_{ab}$, as a function of pulsar antenna response functions and the GWB power distribution on the sky \citep{Taylor_bumpy, Pol_forcast}. The antenna response functions of an Earth-pulsar system are
\begin{equation} \label{eq:response}
    \mathcal{F}^A (\hat{p}_a,\hat{\Omega}) =
    \frac{1}{2} \frac{ \hat{p}_a^i \;\hat{p}_a^j }{ 1-\hat{\Omega}\cdot\hat{p}_a } e^{A}_{ij}(\hat{\Omega})
    ,
\end{equation}
where $\hat{p}_a$ is a unit vector pointing in the direction of pulsar $a$, $\hat{\Omega}$ points to the origin of GWs\footnote{We caution the reader that often $\hat\Omega$ is used as the direction of GW propagation in the literature, rather than the origin vector. Hence the unfamiliar negative sign in the denominator of \autoref{eq:response}. However, all our equations and code is internally consistent with our usage of $\hat\Omega$ as the origin vector.}, $A \in [+,\times]$ denotes the GW polarization, and $e^A_{ij}(\hat{\Omega})$ are the GW polarization basis tensors with $i$ and $j$ corresponding to spatial indices \citep{Taylor_bumpy, Konstandin2025}. Note that while the full response of a pulsar to GW includes an additional factor from the pulsar term, this is uncorrelated between pulsars and (as is commonly done) we neglect it here for simplicity \citep{Pol_forcast, Konstandin2025}.

We can construct the general form of the ORF between distinct pulsars $a$ and $b$ from pairs of these response matrices \cite{Mingarelli2013, TaylorGair2013, Taylor_bumpy}:
\begin{equation}
    \Gamma_{ab} = \int_{s^2} d^2\hat{\Omega} P(\hat{\Omega}) 
    \sum_{A} \left( 
    \mathcal{F}^A(\hat{p}_a,\hat{\Omega}) \mathcal{F}^A(\hat{p}_b,\hat{\Omega})
    \right)
    .
\end{equation}

As is often done, we can turn this integral into a more easily evaluated numerical sum by discretizing our sky into equal-area pixels using HEALPix \citep{Healpix}, while also assuming that different areas of the sky are independent, i.e. $P(\hat\Omega') = \int P(\hat\Omega)\delta^2(\Omega,\Omega')d^2\hat\Omega$. This enables us to simplify our double sky integral into a simpler sum over sky pixels indexed by $k$
\begin{equation} 
    \Gamma_{ab} = \sum_k \frac{3}{2N_{\rm pix}} P_k \left[ 
    \mathcal{F}^+_{a,k} \mathcal{F}^+_{b,k} +
    \mathcal{F}^\times_{a,k} \mathcal{F}^\times_{b,k} 
    \right]
    .
    \label{eq:Gamma_ab}
\end{equation}
where $P_k$ is the angular power from pixel $k$, $N_{\rm pix}$ is the total number of sky pixels, and $\mathcal{F}_{a,k}^A =\mathcal{F}^A(\hat{p}_a,\hat{\Omega}_k)$ \citep{Gair2014, Taylor_bumpy}. Importantly, $P_k$ is normalized such that $\sum_k P_k = N_{\rm pix}$. 

If we were to attempt to jointly constrain the power in all pixels simultaneously, then we must ensure that our system of equations is not under-determined. Assuming all pulsar pairs are equally weighted, the number of sky pixels we model should be fewer than or equal to the number of pulsar pairs \citep{RomanoCornish2017}. For the HEALPix pixelization, which uses an $N_{\rm side}$ parameter to control the number of pixels, we can make an inequality to determine the maximum value
\footnote{Note that while this equation allows for any value, $N_{\rm side}$ is generally a power of 2.}

\begin{equation}
    N_{\rm side} \leq 
    \left[ \frac{N_{\rm psr} (N_{\rm psr}-1)}{24} \right]^{1/2}
    .
\end{equation}

Let us now render \autoref{eq:Gamma_ab} more compact. The ORF model as a vector over all pulsar pairs is  \citep{Pol_forcast}:
\begin{equation} 
    \vec{\Gamma} = \bm{R} \vec{P}
    ,
    \label{eq:anisotropic_orf}
\end{equation}
where $\vec{P}$ is the vector of GWB power values over all sky pixels, and the matrix of correlated responses for a pair of pulsars $(a,b)$ is 
\begin{equation} 
    \bm{R}_{ab,k} = \frac{3}{2N_{\rm pix}} \left[ 
    \mathcal{F}^+_{a,k} \mathcal{F}^+_{a,k} +
    \mathcal{F}^\times_{a,k} \mathcal{F}^\times_{a,k} 
    \right]
    .
    \label{eq:full_response_matrix}
\end{equation}

With these definitions, we now write a chi-squared function and corresponding likelihood function which we will eventually minimize with respect to $\vec{P}$, 
\begin{equation} 
    \chi^{2} = 
    (\vec{\rho} - A_{\rm gw}^{2} \bm{R}\vec{P})^T \bm{C}^{-1} (\vec{\rho} - A_{\rm gw}^{2} \bm{R}\vec{P})
    ,
    \label{eq:anisotropy-chi-square}
\end{equation}
\begin{equation}
    p(\vec{\rho}|\vec{P}) = 
    \frac{1}{\det \left(2\pi\bm{C}\right)}
    \exp \left( -\frac{1}{2} \chi^2\right)
    \label{eq:anis_likelihood}
\end{equation}
where $\bm{C}$ is the pulsar pair covariance matrix that has been approximated as a diagonal matrix of the variances of each pairwise estimator in the weak-signal regime \citep{Gersbach2025},
\begin{equation}
    \bm{C} \approx \bm{C}_{0} \equiv {\rm diag} (\vec{\sigma}_{0}^2)
    ,
    \label{eq:pair_independent_covariance}
\end{equation}
and $\vec{\sigma}_{0}^2$ is the vector of all individual pairwise variances in \autoref{eq:OS_sig_ab}. The efficacy of this approximation will be addressed in \autoref{sec:improve}.

One may then minimize the $\chi^2$ function (or equivalently maximize the likelihood) in \autoref{eq:anisotropy-chi-square} over $\vec{P}$ to find the solution
\begin{equation} 
    \hat{P} = \left( \bm{R}^T \bm{C}_0^{-1} \bm{R} \right)^{-1} \bm{R}^T \bm{C}_0^{-1} \vec{\rho}
    .
    \label{eq:analytic_pixel_solution}
\end{equation}

The problem with this solution is that the Fisher information matrix, $\bm{R}^T \bm{C}_0^{-1} \bm{R}$, is numerically unstable due to the high covariance between pixels. There exists much efforts to condition this matrix \cite{MPTA_anis, Grunthal2026}, however, this paper will instead focus on two more robust approaches: the radiometer basis, and forward modeling using the square-root spherical harmonic model.

\subsection{The Radiometer Basis}

The radiometer basis uses the sky pixelation approach, but instead of solving for the maximum likelihood solution over all pixels, we solve for each pixel independently as if all GWB power were concentrated in a given pixel. This can be written, along with its corresponding variance as
\begin{equation}
    \hat{P}_{k} = 
    \left( \bm{R}_{k}^T \bm{C}_{0}^{-1} \bm{R}_k \right)^{-1}
    \bm{R}_{k}^T \bm{C}_{0}^{-1} \vec{\rho}
    ,
\end{equation}
\begin{equation}
    \sigma_{P_k}^{2} = 
    \left( \bm{R}_{k}^T \bm{C}_{0}^{-1} \bm{R}_k \right)^{-1}
    ,
\end{equation}

where $\bm{R}_k$ is the vector of all pulsar pair responses for pixel $k$\footnote{Many codes, including MAPS, which implement this scheme will use a clever trick to compute all pixel powers and uncertainties simultaneously by zeroing the off-diagonal components of the Fisher matrix, $\bm{R}^T C_{0}^T \bm{R}$, before solving using \autoref{eq:analytic_pixel_solution}.}.
While simple, this solution is far from ideal, as the analysis is effectively fitting a custom ORF for each pixel as if the other pixels do not exist. Despite this drawback, the radiometer basis is suitable for point-like anisotropy searches, such as may be expected from a shot-noise-dominated GW sky \citep{Lamb2024, Sato-Polito2024}. 

One can also compute a per-pixel signal-to-noise ratio (SNR),
\begin{equation} 
    {\rm SNR}_k = \hat{P}_{k} / \sigma_{P_k}
    \label{eq:radiometer_snr}
\end{equation}

However, as with all detection statistics, it should be calibrated against a distribution of the null hypothesis. The choice of null distribution is non-trivial and discussed in \autoref{sec:improve}.

\subsection{Square-root spherical harmonic basis}

The square-root spherical harmonic basis is a variation of the spherical-harmonic decomposition of GWB power, but resolves the problem of non-physical map solutions where negative power can be implied by the fitted spherical harmonic coefficients. The GWB power in each pixel $\vec{P}$ is modeled as
\begin{equation}
    P_{k} = 
    \left[ \sum_{L=0}^{L_{\rm max}} \sum_{M=-L}^{L}  b_{LM} Y_{LM,k} \right]^{2}
    ,
\end{equation}
where $b_{LM}$ are the search coefficients, and $Y_{LM,k}$ are the real-valued spherical harmonics evaluated at pixel $k$. In this equation we want to ensure that we are using sufficient $L_{\rm max}$ which is most often set $L_{\rm max} \approx \sqrt{N_{\rm psr}}$ \citep{Boyle&Pen2012, RomanoCornish2017}. 
However, this choice should not be considered static in robust searches. The resolving power of PTA anisotropy is dependent on a number of factors including the anisotropic distribution of pulsars \citep{NG15_anis} and the non-uniform timing baselines and sensitivities of each pulsar \citep{NG15_dataset, NG15_anis}. A more robust search would involve determining an optimal $L_{\rm max}$ for each analysis though model selection schemes like the Akaike Information Criterion (AIC) or Bayesian Information Criterion (BIC).

By construction, the power in all pixels is positive for any set of $\{b_{lm}\}$. However, since our model is non-linear in these coefficients, and since we want to avoid numerical instability issues by using forward modeling, we now require non-linear minimization of the $\chi^2$ function. This can be accomplished through a python package called \texttt{MAPS}\citep{Pol_forcast}. \texttt{MAPS} makes use of an extensive minimizer package called \texttt{lmfit}\citep{lmfit} which implements several non-linear solving methods. 

With square root spherical harmonics, our full parameter space, controlled by the maximum spherical harmonic $L_{\rm max}$, shrinks significantly to $N_{\rm param} = \left(1+L_{\rm max}/2\right)^2$; making non-linear fitting a viable approach for this basis. This then constitutes forward modeling of anisotropy through the OS framework, rather than the instability-prone inverse solving of \autoref{eq:analytic_pixel_solution}.

\citet{Pol_forcast} also propose a detection statistic for this basis. Called the \textit{anisotropic SNR}, we calculate this through a likelihood ratio
\begin{equation} 
    {\rm SNR} = \sqrt{2 \ln\left( 
    \frac{ p(\vec{\rho}|\vec{P}_{\rm max}) }{ p(\vec{\rho}|\vec{P}=\vec{1}) }
    \right)}
    ,
    \label{eq:sqrt_spherical_snr}
\end{equation}
where $p(\vec{\rho}|\vec{P}_{\rm max})$ is the maximum likelihood value for the anisotropic model and $p(\vec{\rho}|\vec{P}=\vec{1})$ is the isotropic maximum likelihood value. Note that in the isotropic model, we set $L_{\rm max}=0$, such that only the monopole remains. Similar to the radiometer basis, we must calibrate this detection statistic using null distributions.

\subsection{Null distributions}

Since our null hypothesis is an isotropic GWB, generating our null distribution requires many isotropic realizations of our correlated amplitudes, $\vec{\rho}_{\rm null}$, that remain consistent with the measured values of $\vec{\rho}$ and $\sigma_{ab,0}$. In previous works such as \citet{Pol_forcast}, these correlated amplitude are drawn from a Gaussian distribution with means of the HD spatial pair correlations for each pulsar pair and variance determined by the diagonal pulsar-pair covariance matrix
\begin{equation}
    \vec{\rho}_{\rm null} = \mathcal{N}(\vec{\Gamma}^{\rm HD}, \bm{C}_0) 
    .
\end{equation}

The efficacy of this process will be discussed in \autoref{subsec:CV}. We can then substitute $\vec{\rho}$ with $\vec{\rho}_{\rm null}$ in our $\chi^2$ from \autoref{eq:anisotropy-chi-square}, for which we can use the same minimization techniques for the square root spherical harmonic basis to generate a null distribution for the anisotropic SNR for which we can find the percentile of our measured SNR to construct an anisotropic $p$-value. This $p$-value then represents the significance of the anisotropic hypothesis against the isotropic hypothesis.

Applying these null distributions to the radiometer basis is also relatively simple. After substituting our isotropic pair correlations into our $\chi^2$, we simply calculate our new pixel power and pixel uncertainty to generate a null SNR distribution per-pixel. Weighting our measured pixel SNRs against our null distribution gives us per-pixel $p$-values. Each pixel of the radiometer basis will construct a `local $p$-value', labeled as $\tilde{p}$, for which the significance of the measured gravitational wave power is weighed against the significance from an isotropic GWB. 

By themselves, these local $p$-values can only reject isotropy on a pixel-by-pixel basis. To support a `global $p$-value' one must create a statistic which combines either the local $p$-values or SNR distributions together. 
Works such as \citet{Konstandin2025b} have been developing robust methodologies which combine these statistics together to construct a `global $p$-value' though this paper opts to use a simple minimum local $p$-value statistic as it remains consistent with the more grounded anisotropic $p$-value from the square root spherical harmonic analysis. We caution readers to not over-interpret these minimum local $p$-values and instead use it as a comparison diagnostic for this work. Other local-to-global approaches will be discussed in \autoref{sec:conclusion}.

Calibrating our $p$-values and local $p$-values is an important step in the frequentist anisotropy pipeline. The problem with the current method is how it generates pair correlations. As will be detailed in \autoref{subsec:PCOS} and \autoref{subsec:CV}, this method lacks any GWB self noise in the pair covariance matrix $\bm{C}$ and does not include the effects of cosmic variance. These missing effects lead to a vast underestimation in the null distributions, leading to an overestimation in the significance of an anisotropic $p$-value \citep{Konstandin2025}.

\section{Improvements to frequentist anisotropy searches} \label{sec:improve}

PTAs are rapidly approaching the stronger GWB signal regimes required to detect anisotropies. With each new dataset, we will be better able to constrain the GWB and potentially discover its origins through this anisotropic framework. However, to construct a pipeline capable of these tasks we need to fix a few remaining issues. In this section, we aim to resolve three key limitations: $(i)$ we incorporate the pulsar pair covariant optimal statistic (PCOS) to better account for GWB self noise; $(ii)$ we use the per-frequency optimal statistic (PFOS) for spectral characterization, and; $(iii)$ we simulate cosmic variance (CV) in null distributions to better reflect statistical significance of GWB anisotropy.

\subsection{Pair covariant optimal statistic} \label{subsec:PCOS}

The first major improvement comes from recent advancements in the OS that relax the weak-signal regime assumption. In the original formulation of the OS, cross-covariance terms are assumed to be subdominant to the auto-covariance, such that $\expect{\X_b \X_a^T} \approx 0$ \citep{Gersbach2025}. While this assumption has been valid for previous datasets, the most recent 15-year dataset from NANOGrav has shown that this assumption is no longer reasonable \citep{NG15_evidence, NG15_pipeline}. The general formalism for the arbitrary-signal regime simply requires calculating the full pulsar pair covariance matrix,
\begin{equation} 
    \bm{C}_{ab,cd} = 
    \expect{ \rho_{ab} \, \rho_{bc} } - \expect{ \rho_{ab} } \expect{ \rho_{bc} }
    .
    \label{eq:pair_covariance}
\end{equation}

The full details of the rank-reduced formalism for this calculation can be found in Appendix C1 of \citet{Gersbach2025}. This generalization modifies the diagonal and fills the off-diagonal components of the pair covariance matrix to better account for GWB self noise. Several papers have shown that the estimation ability of the PCOS is far more accurate than the original OS in every regime \citep{Gersbach2025, NG15_evidence, NG15_pipeline}. Despite its successes, this correction is still only an approximation as it lacks any HD-deviating effects from cosmic variance, as will be discussed in \autoref{subsec:CV} \cite{Allen_CV, Konstandin2025}.

This generalization requires some assumptions to be made, since the covariance matrix depends on the GWB amplitude. This poses a problem of circularity as the amplitude is the goal of PCOS estimation. The solution used in \citet{Gersbach2025} applies the same logic that the original OS uses when setting the spectral shape, as discussed in \autoref{sec:current}. When employing signal and noise parameter estimates derived from a pilot CURN analysis, we use the CURN amplitude as an estimate for the GWB amplitude in the covariance matrix. This will likely overestimate the total covariance; thanks to the correlated common process always being a subset of the uncorrelated common process. However, in practice, the estimator and its variance appear unbiased in every signal regime tested with simulations \cite{Gersbach2025, NG15_pipeline}.

When applying the PCOS to an anisotropy search, the correct construction of the pulsar pair covariance matrix would involve using the multi-component optimal statistic (MCOS)\cite{Sardesai2023} in which each component would be an ORF for a specific pixel. Borrowing the derivation from Appendix C1 of \citet{Gersbach2025}, the MCOS needs to generalize several scalar quantities from the PCOS. The most important is the expected correlated power, $\expect{\rho_{ab}} = A^2_{\rm gw} \Gamma_{ab}$. For the multi-component case, we can describe our total correlated power as 
\begin{equation}
    \expect{\rho_{ab}} = \bm{R}_{ab} \vec{P}
    ,
\end{equation}
where $\bm{R}_{ab}$ is a pulsar pair slice of the pixel response matrix, $\bm{R}$, from \autoref{eq:full_response_matrix}, and $\vec{P}$ is the vector of the per-pixel powers.

The major problem here is that we do not have an estimate of the per-pixel powers prior to the analyses. With a single ORF, we can assume that the estimated common power from the pilot CURN analysis is due to that ORF. However, when we have multiple pixels, we do not know a priori the power in each pixel before conducting the analyses. 
One may also wish to use an iterative approach to this problem in which we update the pixel powers after solving, then solving again until reaching a stable point where the expected pixel powers and the resulting pixel powers agree. However, in its current state the pair covariance matrix requires roughly a minute to compute a single matrix, multiplying the computational cost dramatically. Additionally, \citet{Gersbach2025} found that a simple iterative scheme for an isotropic pair covariance matrix can lead to numerical instabilities. Instead, for this work, we make an additional, reasonable assumption that the isotropic component of the GWB (the monopolar contribution) dominates the correlated power, such that the ORF is HD to a first approximation.
While suboptimal, this approximation follows from the fact that point sources of GWs with a large number of pulsar pairs will, when averaged over pulsar separation angle, approximate the HD curve \citep{CornishSesana2013}.

With this assumption we can then construct the pair covariance matrix with
\begin{equation}
    \expect{\rho_{ab}} \approx A_{\rm gw}^2 \Gamma_{ab}^{\rm HD}
    ,
\end{equation}
where $\Gamma_{ab}^{\rm HD}$ is the HD correlation for pulsars $a$ and $b$. 

\begin{figure}
    \centering
    \includegraphics[width=\linewidth]{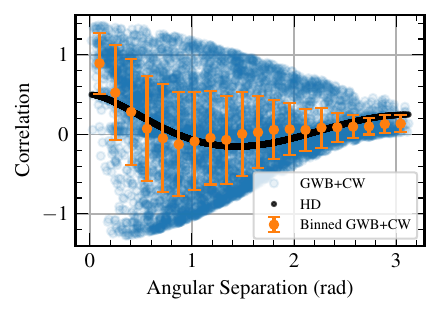}
    \caption{The pulsar pair separation versus correlation plot showing the true pair correlations for a single source containing 90\% of the GW power in a single pixel with the remaining 10\% as an isotropic GWB in blue. Binning these pair correlations by separation angle results in the orange points with the standard deviation represented by the error bars. The binned correlations agree fairly well with the HD solution shown in black, validating our HD covariance matrix assumption.}
    \label{fig:CW_pair_correlations}
\end{figure}

We can test the validity of this statement by by creating a set of pulsar pairs where 90\% of the GW power comes from a single pixel while the remaining GW power is spread evenly among all pixels. (This test mimics the the third frequency of our simulated dataset which will be detailed in \autoref{sec:dataset}.) In \autoref{fig:CW_pair_correlations}, we can see that while the actual pair correlations are spread around the HD curve, by binning pairs correlations by separation bins, the resulting bin-averaged correlation agrees well with the HD curve.

As a final note on pulsar pair covariance, it is not usually used when calculating the SNR of an isotropic GWB, since in that scenario the goal is to estimate the number of null hypothesis (i.e., CURN) standard deviations the signal statistic is from zero \citep{Gersbach2025}. However, for both anisotropy parameter estimation and SNR calculation, we must incorporate pair covariance. The null hypothesis for GWB anisotropy is isotropy, meaning that GWB self noise is present in both models. Hence, pulsar pair covariance will be a necessity for all future anisotropic searches using the optimal statistic.

\subsection{Per-frequency optimal statistic} \label{subsec:PFOS}

The next major improvement is the recently developed per-frequency optimal statistic (PFOS) \cite{Gersbach2025}. The PFOS is a generalization of the OS in which we replace the spectral shape assumption with per-frequency sets of pairwise correlated power estimators and power spectral density (PSD) estimators. The PFOS pairwise estimator and variance are written as
\begin{equation}
    \rho_{ab,n} = \frac
    {\X_a^T \phitilde (f_n) \X_b}
    {\trace{\Z_a \phitilde (f_n) \Z_b \Phi(f_n)}}
    ,
\end{equation}
\begin{equation}
    \sigma_{ab,n,0}^2 = \frac
    {\trace{\Z_a \phitilde (f_n) \Z_b \phitilde (f_n)}}
    {\trace{\Z_a \phitilde (f_n) \Z_b \Phi (f_n)}^2}
    .
\end{equation}
where the $n$ subscript on $\rho_{ab,n}$ indicates the frequency bin the correlations are measuring, and $\phitilde (f_n)$ is a frequency selector matrix that selects the sine and cosine components for frequency $n$ in both the $\X_a$ and $\Z_a$ matrices. The other new matrix, $\Phi (f_n)$, is the estimated shape of the spectrum, which can be written as the estimated PSD normalized at the frequency to be analyzed $\Phi (f_n) \equiv \phi / (S(f_n)\Delta f)$. Like with the original OS, the PFOS uses a preliminary CURN search, often using a variable spectral-index power-law for the spectrum, to inform the shape of the GWB spectrum. The effects of the CURN spectral model choice will be discussed further in \autoref{sec:conclusion}.

The $\chi^2$ statistic from \autoref{eq:anisotropy-chi-square} can then be redefined to be frequency dependent:
\begin{equation} 
    \chi^{2} = 
    (\vec{\rho_n} - S_n \, \bm{R}\vec{P}_n)^T \bm{C}^{-1}_{n} (\vec{\rho}_n - S_n \, \bm{R}\vec{P}_n)
    ,
    \label{eq:pfos_anisotropy-chi-square}
\end{equation}
where $\vec{\rho_n}$ represents the vector of all pairwise estimators $\rho_{ab,n}$, $S_n$ is the PSD within the $n$-th frequency bin, $\vec{P}_n$ is the vector containing all pixel powers for frequency $f_n$, and $\bm{C}_n$ is the pair covariance matrix discussed in \autoref{subsec:PCOS} modified for the per-frequency case, which is defined as
\begin{equation}
    \bm{C}_{ab,cd,n} = 
    \expect{ \rho_{ab,n} \, \rho_{bc,n} } - \expect{ \rho_{ab,n} } \expect{ \rho_{bc,n}}
    ,
\end{equation}
and is derived in \citet{Gersbach2025}. Note that we define $S_n \equiv S(f_n) \Delta f$ as the TOA residual PSD within frequency bin $n$, giving it units of [time]$^2$. 
Note that in this derivation, each frequency of the PFOS is independent of each other. While sub-optimal, we will discuss the validity, and potential improvements to this derivation later in \autoref{sec:conclusion}.

By incorporating the PFOS into our anisotropy pipeline, we gain the ability to dissect both the spectral composition of the GWB through frequency separation and the spatial correlations through the anisotropy formalism. The PFOS enables us to construct GWB maps at each frequency via the radiometer and square-root spherical harmonic bases, but also allows us to calculate per-frequency SNRs. In particular, the square-root spherical harmonic SNR can be calibrated to calculate per-frequency anisotropic $p$-values. 

One subtlety is the ``look-elsewhere effect", which, stated simply, is that by conducting $N$ independent trials, we should not be surprised by a $p$-value below a $1/N$ threshold. This necessitates a correcting factor to compensate. For the square-root spherical harmonic $p$-value, the PFOS tests each frequency independently of others, meaning the number of independent trials is the number of frequencies $N_{\rm freq}$. In this context, the simplest correcting factor is the \textit{Bonferroni correction} \cite{Dunn1961}. This correction changes the threshold value at which the null hypothesis is rejected from $p<\alpha$ into $p<\alpha/N$, or equivalently, we can inflate our measured $p$-values and hold the thresholds constant, $p \times N < \alpha$. 
While we will be showing mostly raw $p$-values, including this correcting factor is a trivial scaling factor.

For the radiometer basis, things are a bit more unclear, since the radiometer basis is used to measure $\gtrsim10^3$ pixels for each frequency. Assuming all of these pixels are completely independent leads to a Bonferroni correction that would overcompensate and eliminate all sensitivity. However, these pixels are far less independent than each frequency, which suggests that the proper correction is somewhere between these two. Similar to the anisotropic $p$-value, we opt not to compensate the local $p$-values. Instead, we use the radiometer basis as a sensitivity map to complement the square-root spherical harmonic basis' anisotropic $p$-value.

\subsection{Cosmic variance null distributions} \label{subsec:CV}

The final improvement we make folds in recent advances in understanding the importance of cosmic variance to anisotropy searches. Cosmic variance \citep[e.g.,][]{Allen_CV, AllenRomano2023, Konstandin2025}, describes how a statistically isotropic GWB will only produce HD cross-correlations in a PTA experiment when averaged over many ensembles, whereas an individual ensemble will exhibit fluctuations about HD due to interference from radiating sources coming from different parts of the sky. The total variance in the OS estimators $\rho_{ab}$ (and $\rho_{ab,k}$) are composed of both pulsar variance and cosmic variance \cite{Allen_CV, AllenRomano2023}. 
\begin{equation}
    \sigma^2_{\rm Total} = \sigma^2_{\rm pulsar} + \sigma^2_{\rm cosmic}
    \label{eq:total_var}
\end{equation}

Pulsar variance arises from the many GWB measurements that can be made for a single pulsar pair separation angle. Two pairs of pulsars with identical separations can still differ in their estimators of the GWB due to the pulsar positions \citep{AllenRomano2023}. Pulsar variance is the dominant effect, and can be reduced by averaging over pulsar pairs with similar separations, called separation bin-averaging. With an infinite number of pulsars, this variance can be completely diminished \citep{Allen_CV}.

Cosmic variance, on the other hand, arises from the single GWB realization to which we have access. The interfering sources of the GWB cause deviations in the estimators that cannot be reduced by bin-averaging, even with an infinite number of pulsars \citep{AllenRomano2023}. If the GWB were statistically isotropic, then cosmic-variance-induced deviations from the expected HD curve could appear falsely as anisotropies \citep{Konstandin2025}. \citet{Konstandin2025} show that neglecting cosmic variance in the frequentist anisotropy pipeline can lead to a $50\%$ false detection rate.

Accounting for cosmic variance in this and any future analysis is of the utmost importance. Unfortunately, it is a non-trivial task to include these effects in the frequentist search pipeline itself. However, \citet{Konstandin2025} employ a clever solution of including its effects in the $p$-value calibration against the null hypothesis. When we generate our isotropic GWB simulations for our null distribution, we can ensure that they contain both pulsar variance and cosmic variance contributions. By doing so, we calibrate our detection statistic against statistical isotropy rather than a GWB realization-averaged isotropy, as was done previously. 

In this section we use a slightly simplified version of equations presented in \citet{Konstandin2025} to generate statistically isotropic pair correlations. The steps we take to simplify can be found in Appendix~\ref{ap:CV}. We first define the full pulsar response function, including pulsar terms, to a GW from sky pixel $k$ within frequency bin $f_n$
\begin{equation}
    R_{a,k,n}^A = \mathcal{F}^A_{a,k} 
    \left[1 - e^
    {-2\pi i f_n L_a \left(1 - \hat{\Omega}_k\cdot\hat{p}_a\right)}
    \right] 
    ,
\end{equation}
where $\mathcal{F}^A_{a,k}$ again represents the pulsar antenna response to a GW originating from sky pixel $k$, and $L_a$ is the distance between pulsar $a$ and Earth. The first term inside the square brackets represents the Earth-term delay, and the second represents the pulsar-term delay. 

Using these response functions, we can now calculate the full pulsar pair correlations,
\begin{equation}
    \hat{\rho}_{ab} = 
    \Re \left[ \sum_n 
    \frac{3}{2N_{\rm pix}} \left[ M_{a,n}^* M_{b,n} \right]
    \times 
    \frac{ S_n }{ \sum_n S_n } 
    \right],
\end{equation}
where $\hat{\rho}_{ab}$ is the dimensionless correlation component\footnote{For a power-law GWB with an amplitude $A_{\rm gw}$, the dimensionless correlation component $\hat{\rho}_{ab} = \rho_{ab} / A^2_{\rm gw}$.}; 
and $n$ indexes frequencies which are assumed to be integer multiples of the inverse PTA time-span, $f_n = n/T_{\rm span}$. The variable $S_n$ is the model PSD of the timing residuals for the GWB at frequency $f_n$. Since this value is normalized over the sum of PSD at all frequencies, we can use the same spectral model the OS uses, which is the diagonal components of $\hat{\phi}$ from \autoref{eq:OS_rho_ab} and \autoref{eq:OS_sig_ab}. Finally, $M_{a,n}$ is the total pulsar response for pulsar $a$ summed over all sky pixels $k$ and polarizations $A$,
\begin{equation}
    M_{a,n} = \sum_{k,A} \hat{h}_{kn}^A R_{a,kn}^A
    .
\end{equation}

The $\hat{h}_{kn}^A$ term represents a GW emitter at each sky pixel $k$, frequency $n$, and GW polarization $A$. As we want to create a realization of a statistically isotropic GWB, we want the amplitudes of both the real and imaginary components of these waves to be zero-mean Gaussian random variables. To achieve this, \citet{Konstandin2025} use a Rayleigh distribution for the amplitudes and a uniform distribution on the phases,
\begin{equation}
    \hat{h}_{kn}^A  = h^A_{kn} e^{i\phi^A_{kn}}
    ,
\end{equation}
where $\phi^A_{kn}~\sim~{\rm U}[0,2\pi]$, and $h^A_{kn}~\sim~{\rm Rayleigh}( \sigma~=~1/\sqrt{2} )$ with a probability density function defined as,
\begin{equation}
    p(x) = 2 x  e^{-x^2} ,
\end{equation}
such that $\langle\hat{h}_{kn}^{A*} \; \hat{h}_{k'n}^{A'} \rangle = \delta_{kk'} \delta_{AA'}$.

We can now create a realization of a statistically isotropic GWB by drawing values for $h_{kn}^A$ and $\phi_{kn}^A$ and computing the pulsar pair correlations. 

Thus far, we have only included signal contributions, without any added noise. This is easily done through noise terms within the pulsar-pair-independent covariance matrix $\bm{C}_{0}$, such that pairwise correlated amplitudes drawn from a statistically isotropic GWB with added noise fluctuations are
\begin{equation} 
    \vec{\rho}_{\rm null} = \mathcal{N} \left(
    \hat{A}_{\rm gw}^2 \vec{\hat{\rho}}_{\rm cv} , 
    \bm{C}_0 \right)
    ,
    \label{eq:rho_null_cv}
\end{equation}
where $\hat{A}^2_{\rm gw}$ is the OS's best-fit GWB amplitude, $\vec{\hat{\rho}}_{\rm cv}$ is the vector of all pulsar pair correlation components, and $\bm{C}_0$ is the pulsar-pair-independent covariance matrix from \autoref{eq:pair_independent_covariance}. While it may seem peculiar to use this versus the pulsar-pair-covariant covariance matrix, we must recall that the pair-covariant covariance matrix partially accounts for the wave interference effects that our generated estimators $\hat{\rho}_{ab}$ include. Hence, we use the pair-independent matrix version to prevent double-counting variance contributions. By generating many realizations of a statistically isotropic GWB and running them through the frequentist anisotropy analysis pipeline, we can accurately calibrate our detection statistic with the presence of cosmic variance. 

If we are using the PFOS, we can use nearly the same scheme, with some further simplifications. Firstly, the calculation of $\hat{\rho}_{ab,n}$ only uses a single frequency $n$, which simplifies our correlations to
\begin{equation}
    \hat{\rho}_{ab,n} = \Re \left[ \frac{3}{2N_{\rm pix}} M^*_{a,n} \, M_{b,n}
    \right]
    .
\end{equation}

We generate frequency-specific noise using the same multivariate normal draw with the PFOS pair-independent covariance matrix $\bm{C}_{n,0}$, 
\begin{equation} 
    \vec{\rho}_{n,{\rm null}} = \mathcal{N}\left( 
    \hat{S}_n \, \vec{\hat{\rho}}_{\rm cv,n} , 
    \bm{C}_{n,0}
    \right)
    ,
    \label{eq:rho_null_cv_pfos}
\end{equation}
where $\hat{S}_n$ is the PFOS estimate of $S_n$, and $\bm{C}_{n,0}$ is the pair-independent covariance matrix for the PFOS in frequency $n$.

Interestingly with the formats presented for the broadband case in \autoref{eq:rho_null_cv}, we can see that every frequency generates a unique realization of cosmic variance, each of which have the same statistical properties. Then the second component does a weighted average based on the PSD in that frequency over all of the frequencies. This agrees with findings from \citet{AllenRomano2025} which found that by averaging the realizations in each frequency, the total cosmic variance decreases. However, this also means that the the single frequency PFOS version in \autoref{eq:rho_null_cv_pfos} contains the maximum cosmic variance.

\section{Simulated Dataset} \label{sec:dataset}

To properly assess these new additions to the frequentist GWB anisotropy detection pipeline, we construct a simulated dataset in which we can control the injected processes. Our dataset was made using the \textsc{Python} pulsar timing simulation package \hyperlink{https://github.com/bencebecsy/pta_replicator}{\textsc{pta replicator}}. This dataset has pulsar timing observations with realistic timing cadences, intrinsic red noise properties, and sky locations. We assume an array of pulsars that mimics the near-future International Pulsar Timing Array Data Release 3 (IPTA-DR3-like). This set of pulsars was constructed in the same manner as in \citet{Petrov2024}, and contains 116 pulsars with 22 years of pulse time-of-arrival (TOA) measurements. 

\begin{figure*}[!t] 
    \centering
    \includegraphics[width=0.58\textwidth]{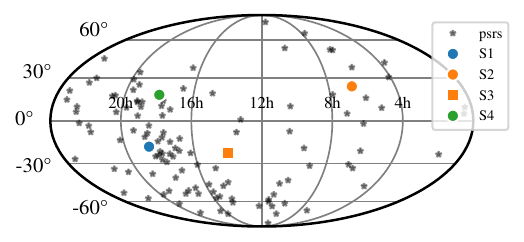}
    \includegraphics[width=0.41\textwidth]{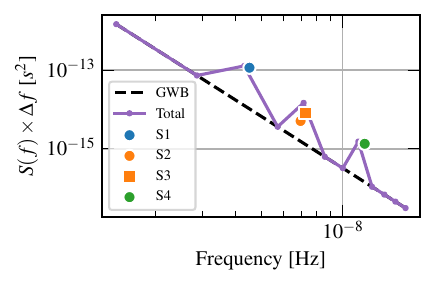}
    \caption{\textit{(Left:)} The sky location of the pulsars (black star) and the CW sources (blue, orange, green) for our test dataset. \textit{(Right:)} The expected total PSD (purple), the expected GWB PSD (black dashed), and the different CW expected PSDs (blue, orange, green) for our test dataset. The total PSD is simply the addition of the GWB with any CW sources within each frequency bin.}
    \label{fig:catered_expected}
\end{figure*}

We start with the $68$ pulsars in the NANOGrav 15-year dataset \citep{NG15_dataset}, then add pulsars from the PPTA DR3 \citep{PPTA_dr3}, EPTA + InPTA DR2 \citep{EPTA_dr2}, and MPTA DR1 \citep{MPTA_gwb} in such a way that we avoid duplicating pulsars. From here, we use the scheme used in \citet{Petrov2024} and developed from \citet{Pol_forcast} to both condense to epoch-averaged TOAs\footnote{Using these epoch-averaged TOAs drastically reduces the total size of the final datasets and eliminates the need for EQUAD and ECORR white noise parameters, however they also tend to underestimate the total contribution of white noise. We apply a $5\times$ inflation to the TOA uncertainties to help combat this.}, and add additional TOAs into the future in a way that either reflects the existing observational cadence or (where such information is lacking) or defaults to a 2-week cadence. Pulsar intrinsic red noise (IRN) properties for these pulsars are taken from measurements within their respective original datasets, and IRN is only injected if IRN was detected in that pulsar. Of the $116$ pulsars, $59$ have IRN, which is injected as a power-law timing-residual power spectral density, where injections span frequencies from $1/T_{\rm span}$ to $30/T_{\rm span}$. The locations of these pulsars are shown as gray stars on the left side of \autoref{fig:catered_expected}.

\begin{table}[!t] 
    \centering
    \begin{tabular}{l|c|c|c|c}
        \hline\hline
        \textbf{Parameter} & \textbf{S1} & \textbf{S2} & \textbf{S3} & \textbf{S4} \\ \hline
        $\mathcal{M}$ $[10^9\,M_{\odot}]$ & $2$ & $2$ & $2$ & $2$ \\ 
        $d_L$ $[\mathrm{Mpc}]$   & 40    & 65    &   130 & 254   \\ 
        $f$ [nHz]   & 4.49  & 6.97 & 7.27 & 12.09 \\ 
        $\Phi_0$ [rad] & 5.22 & 0.90 & 3.45 & 2.50 \\ 
        $\psi$ [rad] & 2.05 & 1.89 & 0.70 & 2.11 \\ 
        $\iota$ [rad] & 0.42 & 1.48 & 0.37 & 2.62 \\ 
        \hline
        $f / f_{0}$ & 3.14  &   4.87    &   5.08    &   8.45    \\ 
        PSD ratio   & 0.90 &  0.35 & 0.56 & 0.88 \\ \hline   
    \end{tabular}
    \caption{The set of binary parameters used to inject 4 CW signals (labeled S1-S4) into our simulated dataset. The initial phase $\Phi_0$, GW polarization $\psi$, and binary inclination $\cos\iota$ were all drawn from uniform random distributions.}
    \label{tab:catered_injection}
\end{table}

We next add GW signals, beginning with an isotropic power-law GWB with an amplitude $A_{\rm gw} = 2.4 \times10^{-15}$, and spectral index $\gamma=13/3$ reference at a frequency of $1/{\rm yr}$, consistent with values measured in the NANOGrav 15-year GWB analysis \citep{NG15_evidence}. On top of this we inject four additional CWs to induce anisotropy in specific frequency bins; binary parameters corresponding to these CWs can be found in \autoref{tab:catered_injection}. These four sources were chosen to be in three different frequency bins: Source 1 (S1) in frequency-bin 3, Source 2 (S2) and Source 3 (S3) in frequency-bin 5, and Source 4 (S4) near the upper edge of frequency-bin 8 (exact values are shown in the row labeled by $f/f_0$ in \autoref{tab:catered_injection}). The expected PSD of each individual source and the GWB are shown on the right side of \autoref{fig:catered_expected}. The source locations were chosen to be somewhat near pulsar sky locations in an attempt to boost the PTA response to anisotropies, shown on the left side of \autoref{fig:catered_expected}.

The initial CW phase, $\Phi_0$, GW polarization $\psi$, and binary inclination $\cos\iota$, were all chosen from uniform random distributions. Binary chirp masses $\mathcal{M}$ were arbitrarily set to $2\times10^9 M_{\odot}$, and distances were selected such that these sources contain roughly $90\%$ of the total GW signal PSD in their respective frequency bins. The exact source-to-total PSD ratios are listed as \textit{PSD ratio} in \autoref{tab:catered_injection}.

\begin{figure}[!t] 
    \centering
    \includegraphics[width=\linewidth]{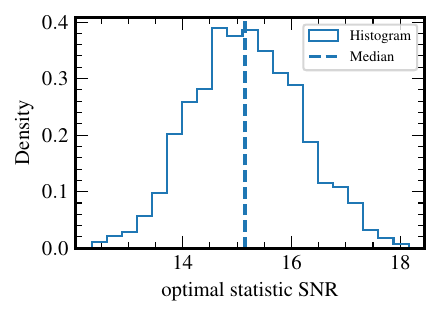}
    \caption{The histogram of the noise marginalized optimal statistic SNR using a $12$-frequency variable spectral-index power-law. The mean SNR is 15.2. }
    \label{fig:os_snr_dist}
\end{figure}

With these injections we find that our optimal statistic isotropic SNR has a median of $15.2$. While loud compared with the NANOGrav 15 year SNR of $5$, this dataset contains nearly double the pulsars, and more than double the TOAs before epoch averaging. The blue histogram and dashed line in \autoref{fig:os_snr_dist} represents the noise marginalized OS SNR distribution and median respectively for our dataset. 

While these injected SMBHB parameter values are not necessarily representative of the first binaries to be detected by PTAs, they were chosen with the intention of exploring the following questions:
\begin{enumerate}
    \item Will the OS or PCOS find a CW present only in one frequency?
    \item Will the PCOS improve anisotropic detection capabilities?
    \item Will the OS or PCOS recover multiple CWs from different frequency bins?
    \item Will the PFOS identify isotropic frequencies?
    \item Will the PFOS find frequencies with CWs in it?
    \item Will the PFOS find the locations of multiple CWs in one frequency?
\end{enumerate}

These questions, which will be returned to in \autoref{sec:discussion}, aim to understand: (i) What improvements PCOS makes to our analysis, (ii) What benefits using the PFOS over a broadband search has, and (iii) What remaining questions we must solve for a robust analysis.

\section{Analysis Implementation} \label{sec:implementation}

The analyses on this dataset use 3 main software tools implemented in \textsc{Python}: \hyperlink{https://github.com/nanograv/enterprise}{\texttt{Enterprise}}, a Bayesian PTA model building package \cite{ENTERPRISE}; \hyperlink{https://github.com/GersbachKa/defiant}{\texttt{Defiant}}, a package to run the various forms of the PTA optimal statistic \cite{Gersbach2025}; and \hyperlink{https://github.com/NihanPol/MAPS}{\texttt{MAPS}}, an implementation of the frequentist PTA anisotropic analysis pipeline \cite{Pol_forcast}. All of these are open-source and can be found on \textsc{GitHub}.

We use \texttt{Enterprise} to construct a CURN model, with the CURN process acting as a spatially-uncorrelated spectral proxy to the GWB. We start with a fixed white noise model with our EFAC for all pulsars set to $1$ without using EQUAD or ECORR. We then model the CURN using a power-law PSD with a varied spectral index and spanning $12$ frequencies from $1/T$ to $12/T$. Intrinsic red noise is modeled for all pulsars with $30$ frequencies. The assumption of a power-law GWB spectrum will lead to some mis-modeling effects in our subsequent PFOS analysis on the true, spiky spectrum. However, in \citet{Gersbach2025} it was found that more agnostic free-spectrum modeling at the initial CURN analysis stage can severely bias the subsequent PFOS due to the large spread in PSD estimates from white noise at high frequencies. A varied spectral index power-law model allows for some deviation from the expected $\gamma=13/3$ spectral index GWB PSD while still bounding high-frequency PSD deviations. This choice will be further discussed in \autoref{sec:conclusion}.

The Bayesian CURN analysis provides posterior samples from which we can construct estimates of the autocovariance matrices in each pulsar $P_a$. We then use these autocovariance estimates within  \texttt{Defiant} to generate the pair-independent OS SNR distribution for our dataset. This distribution, shown in \autoref{fig:os_snr_dist}, has an average SNR of 15.2; this loudness of the GWB signal was found to be necessary in subsequent efforts to find anisotropy by rejecting the isotropy hypothesis.

Using \texttt{Defiant}, we analyze with three different types of OS implementations with successive modeling improvements in order to compare performance. These are the optimal statistic (OS), the pair covariant optimal statistic (PCOS), and the per-frequency optimal statistic (PFOS; which includes pair covariance). These OS analyses provide the following quantities:
\begin{itemize}
    \item \textbf{OS:} $\{\rho_{ab}\}$, $\bm{C}_0$, $\hat{A}^2_{\rm gw}$
    \item \textbf{PCOS:} $\{\rho_{ab}\}$, $\bm{C}_0$,  $\bm{C}$, $\hat{A}^2_{\rm gw}$
    \item \textbf{PFOS:} $\{\rho_{ab,n}\}$, $\{\bm{C}_{n,0}\}$, $\{\bm{C}_n\}$, $\{\hat{S}_n\}$
\end{itemize}

We then pass these quantities to \texttt{MAPS}, in which all operations specific to GWB anisotropy are carried out, including the radiometer and the square-root spherical harmonic analysis. 

For the radiometer basis, we group the pixel amplitude and pixel uncertainty into the pixel SNR from \autoref{eq:radiometer_snr}, as it gives a better understanding of the areas to which the PTA is sensitive and has detectable GWB power. For the square root spherical harmonic basis, \texttt{MAPS} produces both a map of power along with a full sky anisotropic SNR.
\footnote{While normally, the anisotropic maps from the square root spherical harmonic basis are normalized such that $\sum_{k} P_k = N_{\rm pix}$, \texttt{MAPS} also fits an additional full sky scaling parameter, $A$, which modifies the isotropic amplitude (or PSD). For our maps, we opt to show the product the pixel powers and this scale factor such that $\sum_{k} A P_k = A N_{\rm pix}$}

For both bases we must calibrate these SNRs against statistical isotropy. Marginalizing the SNRs over uncertainties in noise processes and cosmic variance involves taking $10^3$ random draws from the initial CURN MCMC chain, one set for each of the analysis methods (OS, PCOS, and PFOS), and, for each draw, creating $10$ realizations of a statistically isotropic GWB using the procedures from \autoref{subsec:CV}. We compute the radiometer per-pixel SNRs and square-root spherical harmonic anisotropic SNR for all of these null (i.e., isotropic) draws. This full distribution properly accounts for spread from both statistical parameter uncertainties and the cosmic variance of an isotropic GWB. These null-hypothesis distributions provide a way for us to compute the $p$-value of the SNR derived from our dataset, giving a calibrated measure of the significance with which we can reject the assumption of isotropy.

Additionally, while we will mostly quote uncorrected (or ``raw'', as we will refer to them) $p$-values, when employing the PFOS method, we should include an additional Bonferroni correction. The simplest form correction is simply $p \times N_{\rm freq}$ for both the radiometer local $p$-values ($\tilde{p}$) and square root spherical harmonic $p$-values. These will be stated in the text where applicable to allow for readers to apply their own corrections. Note that \citet{PostPredict_1} show that the correct form for $p$-value calibration should be the posterior predictive $p$-value, which calibrates each parameter-vector draw from the CURN MCMC chain against a null distribution to form a distribution of $p$-values rather than SNRs. While we agree that this scheme should be used for the analysis of real datasets, the computational expense is dramatically larger, and as such, we opt for the more approximate, lighter approach. 

All analyses were carried out on Vanderbilt University's Advanced Computing Center for Research and Education (ACCRE), where we used approximately $250,000$ CPU hours to run the entire analysis pipeline. While this computing need is large, it remains faster than any comparable methods used previously, and can be massively parallelized, unlike a Bayesian MCMC analysis.

\section{Results} \label{sec:results}

\begin{figure}
    \centering
    \includegraphics[width=\linewidth]{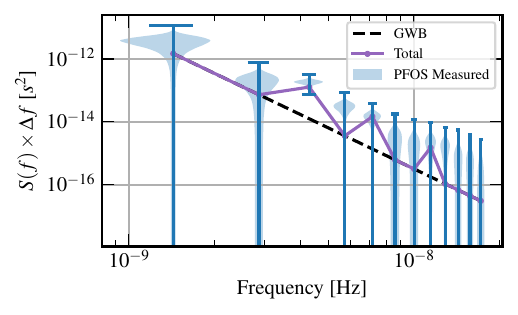}
    \caption{The resulting spectral estimates from the PFOS on our dataset, including uncertainty sampling \citep{Gersbach2025}. The purple represents the injected total PSD in this dataset. The PFOS minorly overestimates many frequencies and majorly overestimates frequency 4 at a percentile of just 0.3\%}.
    \label{fig:catered_PFOS_spectrum}
\end{figure}

In this section, we show the results of applying this new frequentist anisotropy pipeline to the dataset detailed in \autoref{sec:dataset}. After running the $12$-frequency varied spectral index CURN analysis in an MCMC algorithm to generate sufficient posterior samples, we first check the results of the PFOS's spectral estimation. From \autoref{fig:catered_PFOS_spectrum}, we see that the PFOS does find the three different power-law excursions in frequency bins $3$, $5$, and $8$. However, the PFOS is overestimating the PSD in most frequencies. The expected PSDs generally fall within the $1-30$ percentile range of the PFOS PSD distribution with the exception of frequency-bin 4 at the $0.3$ percentile. There are many reasons why the PFOS may struggle with a dataset like this, including poor spectral modeling in the CURN assumptions, frequency-bin covariances (i.e., ``spectral leakage''), and the influence of a strong CW being only approximately HD in its correlation signature. We further discuss possible issues and potential improvements to this recovery in \autoref{sec:discussion} and \autoref{sec:conclusion}.

\subsection{Radiometer Basis}
\subsubsection{Broadband}
We first perform a (broadband) radiometer search using both the OS and PCOS methods. From the $10^3$ individual noise-draw SNR maps from each method, we take the median of each individual pixel SNR distribution to create a single map. These SNR median maps for the OS and PCOS exhibit very consistent behavior, with a hot-spot right where S1 is located, and a roughly even SNR surface elsewhere. This is exactly as one might expect, due to the much larger PSD of S1. However, since the OS lacks the extra accounting for pulsar pair covariance, the smaller values of the pairwise uncertainties artificially inflate the OS's SNRs, making direct comparisons between the SNR maps unhelpful. Therefore, we do not show radiometer SNR maps, but rather attempt to mitigate this behavior by using our statistically isotropic null distributions to calibrate both maps by the expected SNR for an isotropic sky, creating our local $p$-value, $\tilde{p}$, maps.

\begin{figure*} 
    \centering
    \includegraphics[width=0.49\textwidth]{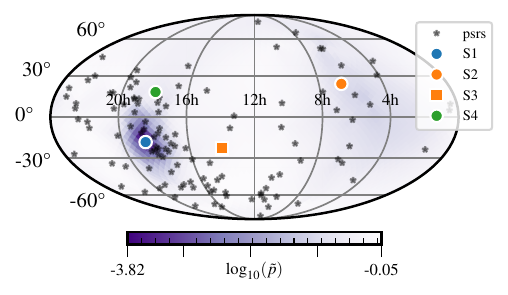}
    \includegraphics[width=0.49\textwidth]{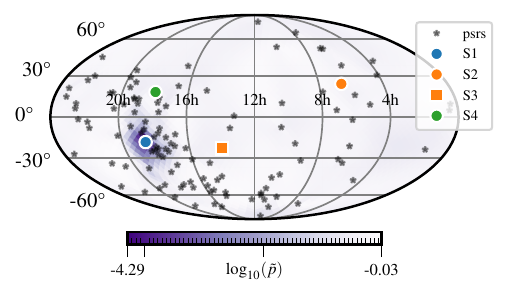}
    \caption{Radiometer local $p$-value maps from the OS and PCOS, respectively, in a broadband analysis. The local $p$-values indicate strong support for the source S1 in frequency 3. The inclusion of pair covariance both improves the minimum local $p$ value from $2 \times 10^{-4}$ to $5 \times 10^{-5}$ and decreases the significant sky region, improving sky localization.}
    \label{fig:catered_radiometer_OS_pval}
\end{figure*}

\autoref{fig:catered_radiometer_OS_pval} shows the radiometer local $p$-value maps for both the OS and PCOS, respectively. They show broad agreement with each other. Both maps have smallest local $p$-values near the loudest source, S1. The OS has a minimum of $2\times 10^{-4}$ while the PCOS has a moderate decrease in the minimum to $5\times 10^{-5}$. Another notable difference is that the sky area with local $p$-value less than $0.01$ is 45 pixels (604~deg$^2$) with the OS, while it is 29 pixels (384 deg$^2$) with the PCOS. This suggests that PCOS is doing a better job at reducing pixel covariances and localizing the sources of anisotropies. 

\subsubsection{Per frequency}
\begin{figure*}
    \centering
    \includegraphics[width=\linewidth]{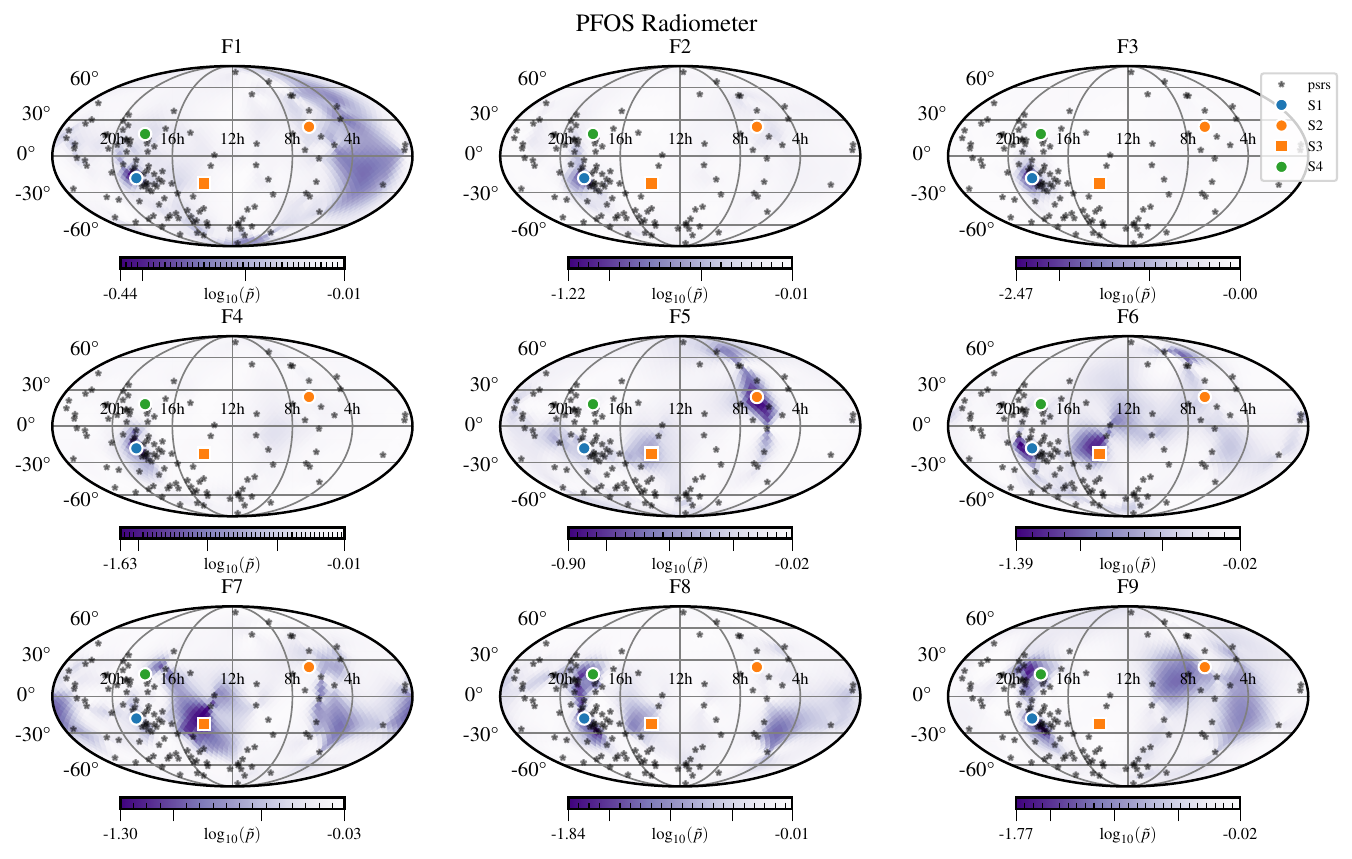}
    \caption{The uncorrected local $p$-values of the radiometer pixel SNR map using the PFOS on the lowest nine frequency bins in our dataset. As a spectral reference, S1 is located in F3, S2 and S3 are in F5, and S4 is located towards the edge of F8 into F9. The smallest local $p$-value in each frequency is represented as the left-hand bound of each figure's color bar. We can see that S1 is strongly indicated in F3 with a minimum local $p$-value of $3 \times 10^{-3}$. S2, S3, and S4 all have some localization though with only minor significance.}
    \label{fig:catered_radiometer_PFOS_1-9}
\end{figure*}

Next, we test the performance of the PFOS. \autoref{fig:catered_radiometer_PFOS_1-9} shows the local $p$-value maps for the lowest nine frequency bins (labeled as F1 through F9). Note that these are raw local $p$-values, not corrected for the look-elsewhere effect. We can immediately see that the anisotropic PFOS recovers frequency-resolved maps of the GWB.

The first frequency bin (F1), containing no GW signals beyond the isotropic GWB, has local $p$-values around $0.5$, with a minimum of $0.36$. F2 however, shows a growing significance region near S1 with a minimum of $0.06$, suggesting spectral leakage may be allowing power to creep backward from F3 into F2.

F3 in particular shows the power of this method. The minimum found was $\tilde{p}=3\times10^{-3}$, far more significant than either F1 or F2. Using the same $\tilde{p}=0.01$ threshold, we find that the enclosed significant sky area is now 6 pixels (80 deg$^2$). However, since the peak is less significant than that of the broadband searches, this threshold region is not directly comparable. Additionally, when using a Bonferroni correction of $p \times N_{\rm freq}$, this minimum, along with the entire region, increases past the $0.01$ threshold level.

In F4 we again see the effects of spectral leakage from S1 at a much stronger level than is seen in F2, likely due to the worse estimation of the PFOS in F4. F5 does exhibit peaks in the local $p$-value near S2 and S3, however, both of these sources are very low significance; S2 has $\tilde{p}=0.19$, and S3 has $\tilde{p}=0.45$. These small differences in local $p$-value are nevertheless interesting, as S3 has a larger expected PSD than S2.

Looking at the minimum local $p$-value as a function of frequency bin, we find that F8 and F9 both show an increase in significance corresponding to S4. Unlike with F3 however, these maps are noisier, and the pixels with the minimum local $p$-value correspond to the group near S1 rather than S4. The remaining 3 frequencies appear similarly to F9 with a grouping near S1. Additionally, one can see the presence of individual pulsar antenna response patterns in almost every frequency, but they are especially apparent in the higher frequencies. This suggests that our sensitivity at higher frequencies may be dominated by a small number of sensitive pulsars rather than the entire array.

\subsection{Square-root Spherical Harmonic Basis}

Our implementation of the square-root spherical harmonic anisotropy analysis requires that we minimize the non-linear $\chi^2$ fitting function over each of $10^3$ parameter vectors drawn from the CURN MCMC chain, thereby marginalizing over uncertainty in PTA noise parameters. This allows us to construct a distribution of maps, from which we summarize with a median map composed of the median power per pixel. The resulting map better accounts for sky regions that are sensitive to variations in the noise parameter values.

\subsubsection{Broadband}

\begin{figure*}[!t]
    \centering
    \includegraphics[width=0.49\textwidth]{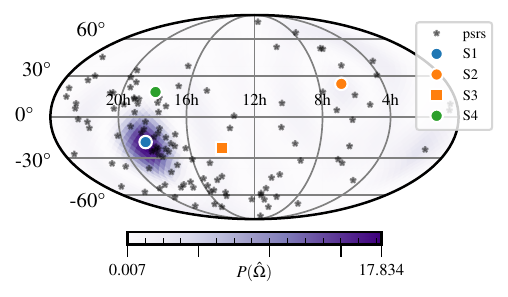}
    \includegraphics[width=0.49\textwidth]{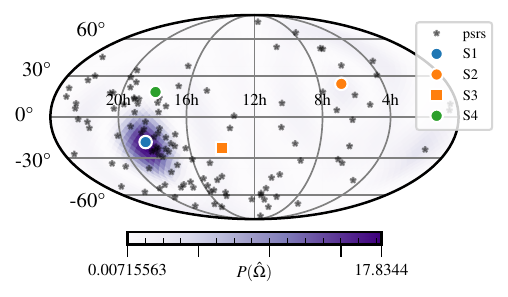}
    \caption{Median pixel maps generated from the square-root spherical harmonic method using the standard OS and PCOS, respectively, on our test dataset in a broadband analysis. Similar to the radiometer analysis, this analysis finds the location of the loudest source, S1, however, both are rather insignificant in their anisotropic $p$-values.}
    \label{fig:catered_sqrt_OS_map}
\end{figure*}

Median maps generated from both the OS and PCOS methods are shown in \autoref{fig:catered_sqrt_OS_map}, exhibiting good consistency. Both methods properly identify the sky location of the loudest source S1; however, upon inspecting their anisotropic $p$-values, we find that the OS and PCOS only give $0.17$ and $0.22$, respectively. This is quite low given the loudness of the injected source. 

\subsubsection{Per frequency}

\begin{figure*}
    \centering
    \includegraphics[width=\linewidth]{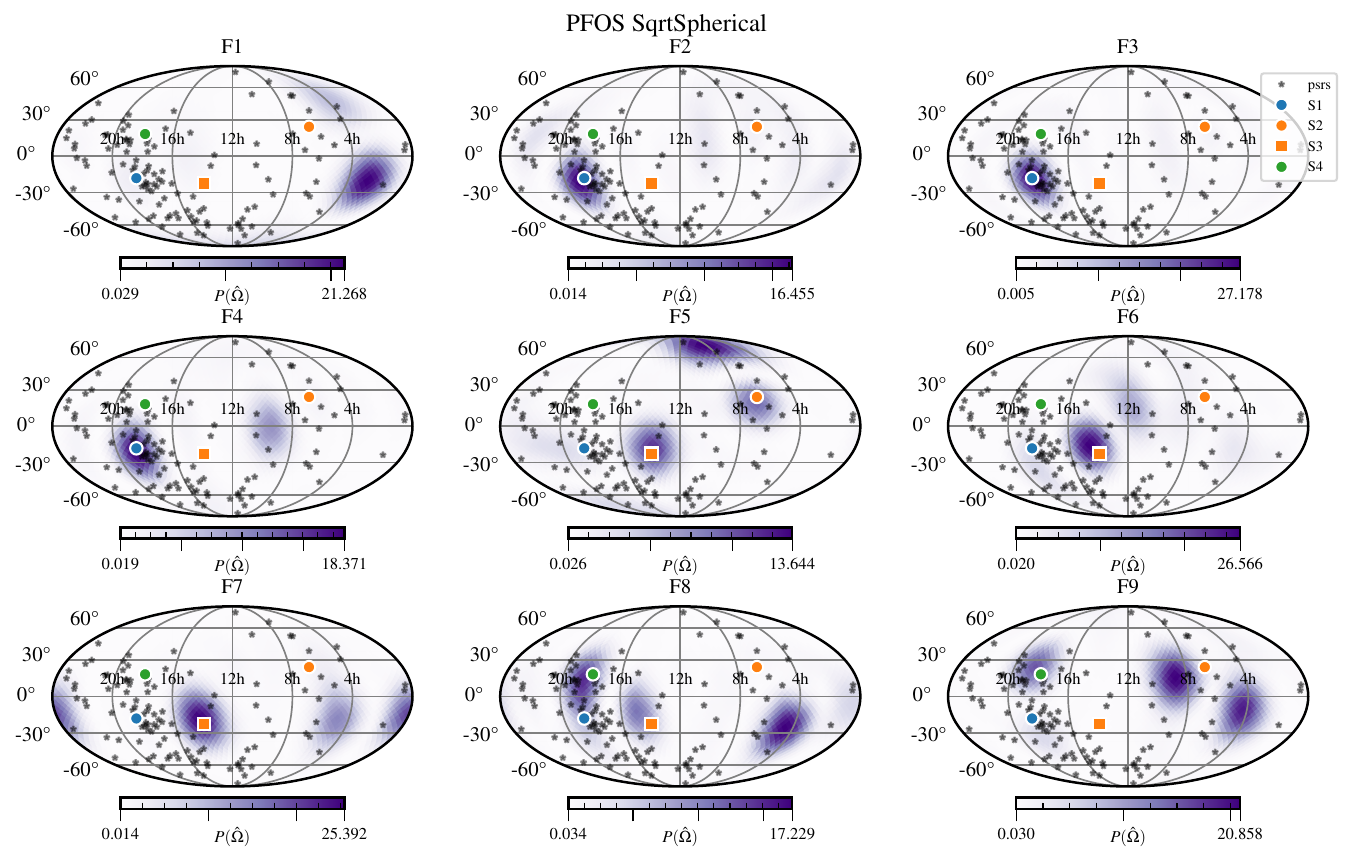}
    \caption{Median-pixel maps for the lowest nine frequency bins in our test dataset, as generated from the square-root spherical harmonic method with PFOS. These maps show the origin of GW power, rather than the direction of GW propagation. All of these maps where created using $L_{\rm max} = 8$, which may not be the optimal solution. These median maps show a strong preference for S1 in frequencies 2-4, F5 shows some preference for S2 and S3 while also showing a sprious hotspot at the north celestial pole, and S4 shows some preference in both F8 and F9. Of these maps, only F3 has anisotropic significance below 1\%.}
    \label{fig:catered_sqrt_PFOS_map}
\end{figure*}

\begin{figure}
    \centering
    \includegraphics[width=\linewidth]{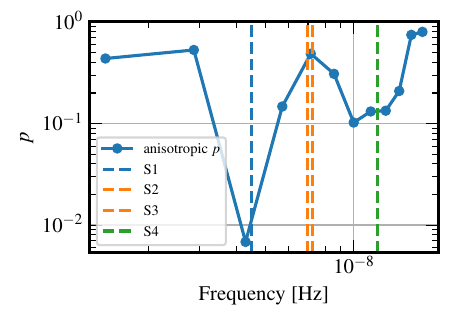}
    \caption{The uncorrected square-root spherical harmonic median SNR $p$-value for the PFOS as a function of the frequency-bin centers. The dashed lines represent the frequencies of the injected individual GW signals. While Only F3 contains significance less than 1\%, we do see a minor bump near S4.}
    \label{fig:catered_sqrt_PFOS_pvalue}
\end{figure}

Median-pixel power maps for the lowest nine frequency bins (F1 through F9) are shown in \autoref{fig:catered_sqrt_PFOS_map}, while the (uncorrected) $p$-values for each frequency bin are shown in \autoref{fig:catered_sqrt_PFOS_pvalue}. F1 and F2 show minimal significance, consistent with the injections. However, F3 shows far greater significance, at an uncorrected level of $6\times10^{-3}$. This frequency, and the corresponding sky location of the hot-spot, coincide with the S1 injection. Similar to the radiometer maps, F4 also identifies the location of S1, corroborating our suspicion of unmodeled frequency-bin covariances. This excess also seems to manifest in the $p$-values as well, reaching $\lesssim 0.2$. 

As expected, the F5 map shows power in the locations of S2 and S3, and the S3 location is recovered with greater power than S2, in qualitative agreement with the injected signal strengths. However, there exists an extra, more powerful hot-spot towards the north celestial pole, a feature which was only weakly seen in the radiometer map. This may be an indicator that multiple sources with similar power can cause interference, such that they manifest false hot spots, or simply just overfitting noise. We discuss this in greater detail in \autoref{sec:discussion}. Regardless, this frequency bin remains well below any detection threshold with an uncorrected $p$-value of $0.5$. 

F6 and F7 exhibit peculiarities, as the $p$-values (shown in \autoref{fig:catered_sqrt_PFOS_pvalue}) have a slight decrease to values of $0.3$ and $0.1$. Moreover, the map has the hottest pixels near S3. This again may be spectral leakage; however, since S3 was not confidently found in F5, this explanation is more challenging to support.

The final source, S4 was injected on the edge of F8, close to F9. The uncorrected $p$-values in these frequencies also show slight decreases to the $0.13$ level for both frequency bins, suggesting that there is indeed some level of anisotropy being identified, though not confidently. Looking at the maps associated with these frequencies, we find hot-spots near S4 in both maps, although neither frequency shows the strongest hot-spots at this location. Rather, they both show an extra hot-spot near S4's antipode, indicating a possible confusion between opposite sky locations. 

\subsection{Frequency correlations}

As an additional quantitative measure of the influence of spectral leakage in our work, we adopt the `r-statistic' from \citet{Moreschi2025}. To use this statistic, we use a spherical harmonic method (such as the square-root spherical harmonic analysis) and compute the auto angular power for each frequency and each mode $l$ as,
\begin{equation}
    C_l^n = \frac{1}{2l+1} \sum_{m=-l}^{l} \left|c_{lm}^n \right|^2
    ,
\end{equation}
where $l$ represents the spherical harmonic mode, $n$ is the frequency index, and $c_{lm}^n$ is the spherical harmonic coefficients. We also need the cross angular power for each pair of frequencies, which is defined as,
\begin{equation}
    Q_l^{nn'} = \frac{1}{2l+1} \sum_{m=-l}^{l} c_{lm}^n \left( c_{lm}^{n'} \right)^*
    .
\end{equation}

We then define the r-statistic as,
\begin{equation}
    \bar{r}^{nn'} = \frac{1}{\sum_{l=0}^{l_{\rm max}} (2l+1)}\sum_{l=0}^{l_{\rm max}} (2l+1) \frac{Q_l^{nn'}}{\sqrt{C_l^n C_l^{n'}}}.
\label{eq:rbar}
\end{equation}
This statistic will be bounded between $-1$ and $1$, where values close to $1$ indicate large, positive frequency correlations between $n$ and $n'$.

\begin{figure}
    \centering
    \includegraphics[width=\linewidth]{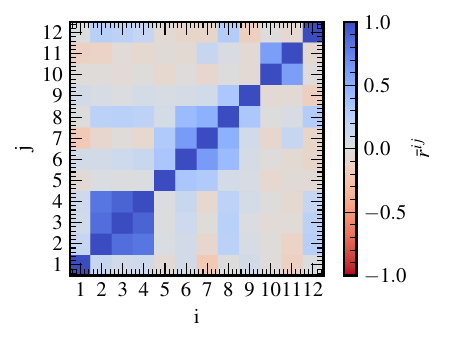}
    \caption{The r-statistic from \autoref{eq:rbar} \citep{Moreschi2025}. This statistic estimates the correlation between frequencies for the spherical harmonic analysis. The correlation between frequency bins 2-4 are very strong indicating spectral leakage is a major problem to solve.}
    \label{fig:rbar}
\end{figure}

We apply this r-statistic to our square root spherical harmonic results by using representative samples of each frequency and computing $\bar{r}^{nn'}$ for all pairs of frequencies. \autoref{fig:rbar} clearly shows the issues this method has with spectral leakage. Under ideal data, the different GW frequencies of our PTA should be independent; however, as can be seen most clearly with frequencies 2-4, the correlation between these frequencies is higher than the other frequency pairs, making it challenging to determine in which frequency bin a source truly belongs.

\section{Discussion} \label{sec:discussion}

We now return to the initial questions we posed in \autoref{sec:dataset}, assessing how well we were able to address them with our new improvements to frequentist PTA anisotropy detection efforts.

\textit{Will the OS or PCOS find a CW present only in
one frequency?---} Both the OS and PCOS were able to find the sky locations of the CW S1 in both the radiometer and square-root spherical harmonic basis, likely thanks to the low frequency and prominent amplitude of the source above the background. However, the anisotropic $p$-values for these broadband methods are lackluster despite its strength. This is an indication that broadband searches for narrowband signals leads to poor detection and parameter estimation prospects for the source. Further testing has also found that the performance of these broadband search methods worsen with sources that have similar relative strengths to the GWB in higher frequency bins. On balance, we advise that broadband search techniques be used in searches where source hypotheses may predict broadband anisotropy.

\textit{Will the PCOS improve anisotropic detection ca-
pabilities?---} The results from using the radiometer basis within the OS and PCOS methods indicate that pair covariance is successful at improving anisotropic detection capabilities. Pair covariance enables further differentiation of the sky pixels in the radiometer basis, tightening the overall hot-spot regions. Additionally, by better accounting for the GWB self-noise, pair covariance more closely matches the total variance from \autoref{eq:total_var}, improving the estimator's ability to differentiate from statistical isotropy. Despite the square-root spherical harmonic basis showing minimal change in the maps between methods, this is mostly due to the locked $L_{\rm max}$ for this basis. Using a variable $L_{\rm max}$, controlled by a model selection scheme, like the Bayesian Information Criterion, may prefer higher modes and smaller localization when modeling pair covariance. 
We also found in \autoref{fig:CW_pair_correlations} that using HD as the assumed correlation signature did not hamper the ability to detect hot-spots of anisotropy.

\textit{Will the OS or PCOS recover multiple CWs from
different frequency bins?---} With our dataset, the OS and PCOS were not capable of finding any additional CWs in different frequency bins. The OS and PCOS maps are dominated by the lowest frequency anisotropic contribution, despite the similar PSD ratios of S1 and S4. The lack of any secondary hot-spot in radiometer maps, even in the locations of these other sources, indicate that the broadband methods are inadequate for a multi-source search in which sources are spread between frequencies. 

\textit{Will the PFOS identify isotropic frequencies?---} The PFOS method shows the clearest improvements of all our new methodological refinements. With it, the radiometer and square-root spherical harmonic analyses both drastically improve on their ability to localize the monochromatic signals of our circular CWs to specific frequencies. Not only this, but the two lowest frequencies, F1 and F2, were correctly identified as isotropic through the assessment of the anisotropic $p$-values of those bins. While F4 did show mildly more significant $p$ values than F1 and F2, it is clear by the maps that frequency covariance (i.e., spectral leakage) are to blame. This means that for particularly loud CWs,---which are more appropriate to model with deterministic templates----their anisotropic contributions in the current approach are not being completely isolated to specific frequency bins. 

\textit{Will the PFOS find frequencies with CWs in it?---} While S1 was admittedly an especially strong source, the PFOS was able to find this source with ease in both bases, finding a far more significant $p$-value than the broadband counterpart. This is a notable improvement from the previous PTA broadband anisotropic pipeline. While S2, S3, and S4 show rather insignificant $p$ values, the maps indicate that these methods are finding the sky locations of these sources, even if they are not confident detections. The bump in $p$ value associated with S4 is also a good sign that the PFOS is able to spectrally separate frequencies enough for multi-frequency, multi-source identification. However, the effects of spectral leakage from S1 likely degrade the detectability of the other sources. Further simulation campaigns with more realistic source populations, without such overwhelmingly prominent sources like S1, may improve multi-frequency, multi-source detectability and localization.

\textit{Will the PFOS find the locations of multiple CWs
in one frequency?---} Finally, the detectability of multiple sources in a single frequency bin is the most challening to assess. The overwhelming strength of S1 likely hinders the separability prospects of S2 and S3. However, from both the radiometer analysis and the square-root spherical harmonic analysis, we see that the PFOS was able to at least find the location of these sources. The downside is the presence of an extra hot-spot towards the north celestial pole, indicating that there may be overfitting or signal interference effects that hinder localization. Further testing on other datasets has found that, in some cases, maps with multiple CWs with similar power contributions in the same frequency bin can cause hot-spots offset from any of the sources, while the anisotropic $p$ value still indicates high detection significance. More testing is required to understand the interactions, confusion, and potential conflation of multiple sources in a single frequency bin, and their impact on the anisotropic $p$ value.

\section{Caveats, Concluding Remarks, \\\& Future Directions} \label{sec:conclusion}

The refinements that we have introduced to the PTA anisotropy detection pipeline are clearly effective. However, there are several caveats to our analyses which should be examined more closely. The first, and likely most important, is in the initial CURN estimation stage. While we chose to use a power-law PSD for our Bayesian CURN model, the true PSD of our overall signal is a poor match to a power-law. Indeed, while the PFOS does attempt to correct for these estimation issues through the more agnostic search, the poor percentiles from the PFOS PSD estimation undoubtedly propagate biases further into the anisotropic stages of the pipeline. While \citet{Gersbach2025} mentions the poor performance when employing a free-spectrum CURN model for the initial spectral estimate, other more informed models, such as a $t$-process \citep{2024ApJ...976..212S} or a spike-slab model, 
have been more successful in preliminary testing.
We recommend others who intend to build upon this work to explore these more generalized Bayesian models, enabling spectral excursions without the unbounded high frequencies of free-spectrum models.

Another limitation of this new pipeline is in the PFOS itself. As constructed, the PFOS models independent frequencies, such that at each frequency we fit a GWB PSD estimate without incorporating information from other frequencies \citep{Gersbach2025}. This scheme, while straightforward, causes issues for sources that may straddle different frequency bins, resulting in spectral leakage and overestimation. Indeed, we see this with our dataset in the PFOS estimations of F4, which saw a notable overestimate of the PSD . 
This propagated into the anisotropy analysis, in which the r-statistic showed measurable leakage of S1 into the adjacent frequencies. While the $p$-value for these frequencies remained weak, the correlations between the sky maps show that poor modeling can artificially inflate significance by leaking into adjacent frequencies. While the inclusion of a more robust Bayesian modeling approach will likely improve beyond what we have shown here, the most robust solution will need to involve fitting the PFOS to all frequencies simultaneously while accounting for frequency covariances similar to the newest Bayesian methods \citep{2025arXiv250613866C}.

While this work assumed a fixed maximum multipole for the spherical harmonic representation of GW power distributions, the ideal scenario would be to instead let the data inform this through model selection. This would aid in preventing model overfitting and underfitting. Building a more robust and data-informed method for determining model hyperparameters, like this maximum multipole, during analyses should be prioritized when applying these methods to real data to ensure the optimal solutions are found.

Methods for constructing the radiometer basis' global $p$-value also requires further study. While we opted to use a basic minimum local $p$-value estimate, there are more mathematically rigorous methods to compute a global $p$-value. One way, analogous to the posterior predictive $p$-value from \citet{PostPredict_1}, would be to construct a distribution of local $p$-values per pixel by taking each set of parameter estimates from the CURN MCMC and creating a set of local $p$-values for each. After repeating this process for many parameter estimates to account for noise marginalization, the resulting minimum $p$-values more accurately represent the excess detectability of the radiometer basis. The downside to this scheme is the computational cost, as estimating $p$-values less than $1/N$ require at least $N$ null draws for each set of parameter estimates, necessitating a $\sim100\times$ increase in the computational resources compared with the methods used in this work. 

Another issue with the radiometer basis are the frequent appearance of one or more pulsar antenna response patterns. This is seen most clearly in the PFOS, showing the characteristic quadrupolar response patterns of different pulsars at different frequencies. These are likely due to some pulsars being more sensitive to particular frequencies than others, arising from noise realizations, timing cadences, and observational time-spans. Because of this, the PFOS methods for both radiometer and square-root spherical harmonic bases may benefit from pulsar dropout-like analyses from \citet{NG11_CW} or \citet{NG15_remove_1b1}, ensuring that a single pulsar does not dominate these map reconstructions.

Another avenue we have begun to explore is the use of principal bases. Since the sensitivity of a PTA is very non-uniform, using generalized bases like the pixel basis are numerically challenging as pixel covariances make analytic solutions deeply challenging and require significant regularization techniques to tame \citep{MPTA_anis, Grunthal2026}. Another similar line of reasoning is to change the basis functions themselves to be linearly independent pixel combinations, using processes like Eigenvalue decomposition for a Fisher information matrix analysis or singular value decomposition for a more general approach. These alternative analyses are proving promising while also reducing the computational complexity needed to analyze these problems. We will explore these in future work.

Finally, we should acknowledge that, despite the strong signals we injected, the significance level of anisotropy that we identified was still rather weak. Undoubtedly, it would be more appropriate to search for these sources, especially ones like S1, using deterministic signal models in a transdimensional multi-source search.


In summary, we find that modeling pulsar pair covariance in the optimal statistic (OS) greatly improves the ability to localize anisotropies in frequentist PTA searches. The benefits of pair covariance go even further when combined with the per-frequency OS (PFOS). When properly accounting for cosmic variance, the PFOS is far better equipped to deal with frequency-specific anisotropies, improving $p$-value significance from a meager $0.2$ level to a more confident sub-$0.01$ level. The PFOS enables our new frequency-specific anisotropic methods, while still remaining computationally feasible thanks to its trivial parallelizability that contrasts with sequential Bayesian MCMC analyses.  

These improvements can also be more widely applied to other anisotropy search techniques, like full-sky pixelated map recoveries, recovery with the linear spherical harmonic basis, eigen-map reconstructions, or any other future schemes that leverage the optimal statistic for their correlation information. Along with the methods themselves, the simplified forms that we have presented for cosmic-variance-limited null distributions are easily calculated, and enable more accurate $p$-value calculations that allow us to robustly distinguish between anisotropy and statistical isotropy.

In the future, we plan to further explore our new frequentist PTA anisotropy framework's performance and subtleties on more realistic GWB signals, constructed from synthetic populations of supermassive black-hole binaries. Effects such as the anomalous hot-spots in multi-source frequency bins, and the potentially-related effects of spectral leakage, are the primary limitations to be addressed when moving to more realistic datasets. Our preliminary tests on such realistic GWB signals, with multiple loud binaries competing in a given frequency bin, have found excellent efficacy in terms of anisotropic $p$ values. However, we have also seen that in some scenarios, these high-confidence maps may miss the sky locations of the CWs, suggesting that source interference may be yet another important effect that has not yet been considered.

\begin{acknowledgments}

We thank Taha Moursy for finding a subtle bug in some of our code. We appreciate the support of the NANOGrav NSF Physics Frontier Center awards \#2020265 and \#1430284. SRT acknowledges support from NSF AST-2307719. KAG and SRT acknowledges support from NSF CAREER \#2146016. SRT is also grateful for support from a Vanderbilt University Chancellor's Faculty Fellowship. AM and AL acknowledge support by the Deutsche Forschungsgemeinschaft under Germany’s Excellence Strategy - EXC 2121 Quantum Universe - 390833306. This work was conducted in part using the resources of the Advanced Computing Center for Research and Education (ACCRE) at Vanderbilt University, Nashville, TN.

\end{acknowledgments}

\nocite{*}

\bibliography{NGFreqAni}

\appendix

\section{Cosmic variance calculation} \label{ap:CV}

In this paper, we employ a slightly simplified formulation for cosmic variance. We start with Equation (7) from \citet{Konstandin2025}, including the normalization factor $N_{ab}$,
\begin{widetext}
\begin{equation}
\begin{aligned}
\rho_{ab} = N_{ab} \!\!\!\!\!\!\sum_{nn',kk',AA'} \left[ 
\tilde{h}_{kn}^{A*} \tilde{h}_{k'n'}^{A'} R_{a,kn}^{A*} R_{b,k'n'}^{A'} \: {\rm sinc}(\pi(f_n -f_n')T) -
\tilde{h}_{kn}^{A} \tilde{h}_{k'n'}^{A'} R_{a,kn}^{A} R_{b,k'n'}^{A'} \: {\rm sinc}(\pi(f_n +f_n')T)
\right]
\frac{\Delta \hat{\Omega}^2 \Delta f^2}{f_n f_{n'}}
+ \\ {\rm c.c.},
\end{aligned}    
\end{equation}
\end{widetext}
where the random compelx variable, $\tilde{h}_{kn}^{A}$, represents the GW at a particular sky pixel $k$, frequency bin $n$, and polarization $A \in [+,\times]$; $R_{a,kn}^{A}$ is the full pulsar response function, including the pulsar term; $T$ is the total observational time span of the PTA; $\Delta \hat{\Omega}$ is the solid angle area of the sky pixel; and $\Delta f$ is the frequency bin width, which is assumed to be the same for all bins.

The first step in simplifying this equation is to define our frequency bins $f_n$. We use the same frequency resolution bins as in the analysis in the main section of the paper, i.e.,
\begin{equation}
    f_n = \frac{n}{T},\quad n= 1, 2, ...,N_{\rm freq}. 
\end{equation}
This definition simplifies the sinc terms, such that
\begin{equation}
\begin{aligned}
{\rm sinc}[\pi(n - n')] &= \delta_{nn'},
\\
{\rm sinc}[\pi(n + n')] &= 0,
\end{aligned}
\end{equation}
which in turn removes one of the sums over frequency,
\begin{equation}
\rho_{ab} = 2 \Re \left\{ 
    N_{ab} \!\!\!\!\sum_{n,kk',AA'} \!\!
    \left[ 
        \tilde{h}_{kn}^{A*} \tilde{h}_{k'n}^{A'} R_{a,kn}^{A*} R_{b,k'n}^{A'} 
    \right] \frac{\Delta \hat{\Omega}^2 \Delta f^2}{f^2_n} 
\right\}
.
\end{equation}
where we have replaced the sum of the term with its complex conjugate by just doubling the real part. Next, we redefine the complex waveforms by using the properties of their Rayleigh distributed amplitudes,
\begin{align}
    \tilde{h}_{kn}^{A} &= 
    \sqrt{\frac{H_n}{\Delta \hat{\Omega} \Delta f }} \;
    \hat{h}_{kn}^{A} \, ;
    \\
    \hat{h}_{kn}^A  &= h^A_{kn} e^{i\phi^A_{kn}}
    ,
\end{align}
where $\phi^A_{kn}$ is drawn from a uniform distribution, ${\rm U}[0,2\pi]$, $h^A_{kn}$ is drawn from a Rayleigh distribution, ${\rm Rayleigh}( \sigma~=~1/\sqrt{2} )$ with a scale factor $\sigma$, and $H_n$ is the PSD of the Fourier modes of the GWB. By making this redefinition of our random complex waves into $\hat{h}_{kn}^A$, we can remove the GWB dependence and simplify the implementation. The choice of scale for the redefined Rayleigh distribution was made such that the expectation results in the simple expression
\begin{equation} \label{eq:expect_hhat}
    \expect{ \hat{h}_{kn}^{A*} \; \hat{h}_{k'n}^{A'}}  =
    \delta_{kk'} \delta_{AA'}
    .
\end{equation}

Substituting the redefined complex waves into the expression for $\rho_{ab}$ gives
\begin{equation}
\rho_{ab} = \Re \left\{ N_{ab} \!\!\!\!\sum_{n,kk',AA'}\!\! \left[ 
\hat{h}_{kn}^{A*} \hat{h}_{k'n}^{A'} R_{a,kn}^{A*} R_{b,k'n}^{A'}
\right] \frac{2 H_n \Delta \hat{\Omega} \Delta f}{f^2_n}
\right\}
.
\end{equation}

We now need to apply an appropriate transformation from $H_n$ to the PSD of the TOA residuals, $S_n$. We can do so through $S_n \propto H_n \Delta f / f^2$; the overall normalization, $N_{ab}$ will be computed later. Using this, and the substitution $\Delta \hat{\Omega} = 4\pi/N_{\rm pix}$, we can write
\begin{equation}
\rho_{ab} = \Re \left\{ N_{ab} \!\!\!\!\sum_{n,kk',AA'}\!\! \left[ 
\hat{h}_{kn}^{A*} \hat{h}_{k'n}^{A'} R_{a,kn}^{A*} R_{b,k'n}^{A'}
\right] 
\frac{8 \pi S_n}{N_{\rm pix}}
\right\}
.
\end{equation}

As a final step before solving the normalization, we introduce an extra factor whose utility will soon become clear:
\begin{equation}
\rho_{ab} = \Re \left\{ N_{ab} \!\!\!\!\sum_{n,kk',AA'} \!\!\frac{3}{2N_{\rm pix}} \left[ 
\hat{h}_{kn}^{A*} R_{a,kn}^{A*}
\hat{h}_{k'n}^{A'}  R_{b,k'n}^{A'}
\right] 
\frac{16 \pi S_n}{3}
\right\}
.
\end{equation}

We solve for the normalization factor $N_{ab}$ by taking the expectation of $\rho_{ab}$, for which we should expect $\expect{ \rho_{ab} } = A^2_{\rm gw} \Gamma_{ab}$. Upon substituting \autoref{eq:expect_hhat}, we find
\begin{equation}
A^2_{\rm gw} \Gamma_{ab} =
\Re \left\{ N_{ab} \!\!\!\!\sum_{n,k,A} \!\!\frac{3}{2N_{\rm pix}} \left[ 
R_{a,kn}^{A*} R_{b,kn}^{A}
\right] 
\frac{16 \pi S_n}{3}
\right\}
.
\end{equation}

From here, we make the (good) approximation that pulsar terms will be uncorrelated between pulsars. This allows us to approximate $R_{a,kn}^{A} \approx \mathcal{F}_{a,k}^A$. Note that this eliminates any remaining complex numbers as well as the frequency dependence of the pulsar response functions, such that
\begin{equation}
A^2_{\rm gw} \Gamma_{ab} =
N_{ab} \sum_{n,k,A} \frac{3}{2N_{\rm pix}} \left[ 
\mathcal{F}_{a,k}^{A} \mathcal{F}_{b,k}^{A}
\right]
\frac{16 \pi S_n}{3}
.
\end{equation}

We can now write our sum of products as a product of sums, while also explicitly summing over the polarization contributions, such that
\begin{equation}
A^2_{\rm gw} \Gamma_{ab} = 
N_{ab} \sum_{k} \frac{3}{2N_{\rm pix}} \left[ 
\mathcal{F}_{a,k}^{+} \mathcal{F}_{b,k}^{+} + 
\mathcal{F}_{a,k}^{\times} \mathcal{F}_{b,k}^{\times}
\right]
\sum_{n}\frac{16 \pi S_n}{3}.
\end{equation}

Notice that the first sum is identical to the definition of our correlation, $\Gamma_{ab}$, in \autoref{eq:Gamma_ab}. This lets us simplify, such that
\begin{equation}
A^2_{\rm gw} \Gamma_{ab} = 
N_{ab} \, \Gamma_{ab}  
\frac{16 \pi}{3}\sum_{n} S_n
,
\end{equation}
whereupon we find the normalization 
\begin{equation}
N_{ab} = 
A^2_{\rm gw} \left( \frac{16 \pi}{3} \sum_{n} 
S_n \right)^{-1}
.
\end{equation}

Placing this back into our full equation, and canceling constant terms that appear in the function and the inverse in the normalization, gives
\begin{equation}
\frac{\rho_{ab}}{A_{\rm gw}^2} = \Re \left\{ \sum_{n,kk',AA'}\!\! \frac{3}{2N_{\rm pix}} \left[ 
\hat{h}_{kn}^{A*} R_{a,kn}^{A*}
\hat{h}_{k'n}^{A'}  R_{b,k'n}^{A'}
\right]
\frac{ S_n }{ \sum_n S_n } 
\right\}
.
\end{equation}

There is one final, key insight to transform this equation into the one used in \autoref{subsec:CV}. This is to identify that the pixel sums only apply to the first term, and that the $k$ and $A$ sums can be separated as a product of sums. If we define
\begin{equation}
M_{a,n} = \sum_{k,A} \hat{h}_{kn}^{A} R_{a,kn}^{A}
,
\end{equation}
then we can place this into our equation's final form
\begin{equation}
\frac{\rho_{ab}}{A_{\rm gw}^2} = 
\Re \left\{
\sum_n 
\underbrace{ \frac{3}{2N_{\rm pix}} \left[ M_{a,n}^* M_{b,n} \right]
}_{\text{Correlation Component}}
\,\,\,\times \!\!\!\!\!
\underbrace{ \frac{ S_n }{ \sum_n S_n } 
}_{\text{Frequency weighting}}
\right\}
.
\end{equation}

This final equation has clear distinctions between the correlation component, which has a unique but statistically identical set of correlations for each frequency, and the frequency weighting, which weights each frequency relative to the PSD strength.

\end{document}